\documentclass[fleqn,10pt]{wlscirep}
\usepackage[utf8]{inputenc}
\usepackage[T1]{fontenc}

\usepackage[flushleft]{threeparttable}
\usepackage{adjustbox}

\usepackage{subfigure}
\usepackage{color}

\usepackage{aas_macros}

\usepackage{amsmath}
\usepackage{amsfonts}
\usepackage{amssymb}
\usepackage{graphicx}
\usepackage{csquotes}

\usepackage{lineno}
\usepackage{setspace}
\usepackage{xcolor}
\usepackage{float}

\def\fermi{{{\em Fermi}-LAT}}

\newcommand{\appropto}{\mathrel{\vcenter{
  \offinterlineskip\halign{\hfil$##$\cr
    \propto\cr\noalign{\kern2pt}\sim\cr\noalign{\kern-2pt}}}}}

\newcommand{\aref}[1]{\hyperref[#1]{Appendix~\ref{#1}}}

\usepackage[backend=biber, style=nature, eprint=false, doi=false]{biblatex}

\usepackage{jabbrv}

\AtEveryBibitem{
  \clearfield{month}
}

\DeclareFieldFormat{journal}{ \itshape #1 }
\DefineBibliographyStrings{english}{andothers = {\normalfont et\addabbrvspace al\adddot}}

\makeatletter
\AtEveryCitekey{%
  \ifcsundef{blx@entry@refsegment@\the\c@refsection @\thefield{entrykey}}
    {\csnumgdef{blx@entry@refsegment@\the\c@refsection @\thefield{entrykey}}{\the\c@refsegment}}
    {}}
    
\defbibcheck{onlynew}{%
  \ifnumless{0\csuse{blx@entry@refsegment@\the\c@refsection @\thefield{entrykey}}}{\the\c@refsegment}
    {\skipentry}
    {}}
\makeatother

\title{Millisecond Pulsars from Accretion Induced Collapse as the Origin of the Galactic Centre Gamma-ray Excess Signal}

\author[1]{Anuj Gautam}
\author[1,*]{Roland M.~Crocker}
\author[2]{Lilia Ferrario}
\author[3]{Ashley J.~Ruiter}
\author[4]{Harrison Ploeg}
\author[4]{Chris Gordon}
\author[5,6]{Oscar Macias}

\affil[1]{Research School of Astronomy and Astrophysics, Australian National University, Canberra 2611, A.C.T., Australia}
\affil[2]{Mathematical Sciences Institute, The Australian National University, Canberra, ACT 2601,  Australia}
\affil[3]{School of Science, University of New South Wales Canberra, The Australian Defence Force Academy, 2600 ACT, Canberra, Australia}
\affil[4]{School of Physical and Chemical Sciences, University of Canterbury, Christchurch, New Zealand}
\affil[5]{Kavli Institute for the Physics and Mathematics of the Universe, University of Tokyo, Kashiwa, Chiba 277-8583, Japan}
\affil[6]{GRAPPA Institute, University of Amsterdam, 1098 XH Amsterdam, The Netherlands}

\affil[*]{Corresponding author: roland.crocker@anu.edu.au}

\newrefsegment

\begin{abstract}
Gamma-ray data from the {\it Fermi}-Large Area Telescope reveal an unexplained, apparently diffuse, signal from the Galactic bulge
\cite{Hooper2011, Gordon2013, Abazajian2014} that peaks near ${\sim} 2$ GeV with an approximately spherical \cite{Daylan2016} intensity profile $\propto r^{-2.4}$ \cite{Abazajian2014, Calore2015} that extends to angular radial scales of at least ${\sim} 10^\circ$, possibly
 to ${\sim} 20^\circ$ \cite{Hooper2013b,Ackermann2017}. 
The origin of this ``Galactic Centre Excess'' (GCE) has been debated with proposed sources prominently including self-annihilating dark matter \cite{Hooper2011,Daylan2016} and a hitherto undetected population of millisecond pulsars (MSPs) %\cite{Abazajian2011,Gordon2013,Macias2018,Bartels2018,Macias2019,Calore2021}.
\cite{Abazajian2011}.
However, the conventional channel for the generation of MSPs has been found  to predict too many low mass X-ray binary (LMXB) systems %\cite{Cholis2015,Haggard2017}
\cite{Haggard2017}
and, because of the expected large natal kicks, may not accommodate \cite{Ploeg2021} the close spatial
correspondence \cite{Macias2018,Bartels2018,Macias2019} between the GCE signal and stars in the bulge.
Here we report a binary population synthesis forward model that demonstrates that
an MSP population arising from the accretion induced collapse (AIC)
of O-Ne white dwarfs in Galactic bulge binaries can naturally reproduce the morphology, spectral shape, and intensity of the GCE signal while also obeying LMXB constraints.
Synchrotron emission from MSP-launched cosmic ray electrons and positrons may simultaneously explain
the mysterious, microwave ``haze''  \cite{Ade2013} from the inner Galaxy.
\end{abstract}
\begin{document}

\flushbottom
\maketitle
\thispagestyle{empty}

\noindent
The spectrum of the GCE resembles those of individual $\gamma$-ray MSPs and also those of globular clusters whose $\gamma$-ray signals are due to large MSP populations% \cite{Chen1991,Abdo2010,Abazajian2011,Gordon2013}
\cite{Chen1991,Abdo2010}. 
While 
there is no secure detection of 
individual millisecond pulsars (MSPs)
in the Galactic bulge or centre,
such a population is not unanticipated \cite{Wang2005}.
One potential discriminant between DM and MSPs as the origin of the GCE is that, while the former should be characterised by
``true'' diffuse emission, the latter is, at least partially, 
expected to arise from the combined emission of many discrete, but dim MSP sources.
Statistical methods 
of {\it point source inference} %\cite{Collin2021}
might distinguish between these two types of emission %\cite{Bartels2016,Lee2016} 
but there is on-going controversy about the reliability of these methods 
given large foreground model uncertainties 
%\cite[e.g.,][]{Leane2019,Leane2020,Leane2020b,Buschmann2020,Chang2020}.
[e.g.,\cite{Buschmann2020}].

A completely independent  argument
for a stellar origin of the GCE is that  
its morphology traces the stars of the inner Galaxy better %\cite{Macias2018,Bartels2018,Macias2019,Calore2021} 
\cite{Macias2018,Bartels2018,Macias2019}
than it traces the radially symmetric distribution expected for dark matter.

An argument advanced against an MSP GCE is that the implied $\gamma$-ray luminosity distribution would be incompatible
with the local and resolved MSP disk population since this predicts too many bright  MSPs \cite{Hooper2016}, though note that this argument is disputed \cite{Ploeg2020}.
Regardless, the bulge MSP population need not share the characteristics of the local Galactic disk MSP population: star formation ceased $\sim8-10$\,Gyrs ago in the bulge %\cite{Nataf2016, Crocker2017} 
\cite{Nataf2016} 
while it is ongoing in the disk.
Because MSPs spin down over time thus becoming less $\gamma$-ray luminous -- with those initially brightest evolving fastest --
the systematic age difference between bulge and disk MSP populations should lead to a relative deficit of bright MSPs in the bulge.

To date, much of the discussion surrounding the tenability of the MSP explanation for the GCE has implicitly adopted the canonical ``recycling'' formation scenario whereby an old, slowly rotating neutron star (NS) formed in a core-collapse supernova (CCSN) accretes material from a companion during Low-Mass X-Ray Binary (LMXB) phases \cite{Radhakrishnan1982}. 
This spins up the NS to millisecond periods enabling it to emit  $\gamma$ radiation. %\cite{Bhattacharya1991}. 
However, a recycling origin for the MSPs predicts that the bulge should host ${\sim} 10^3$ LMXBs, while observations by INTEGRAL have uncovered only 42 LMXBs with an additional 46 candidates within a $10^\circ$ radius of the Galactic Centre \cite{Haggard2017}. 

This argument, however, neglects the existence of a separate mechanism which may produce $\gtrsim$ 50\% of all observed MSPs %\cite{Ferrario2007,Hurley2010,Tauris2013,Ruiter2019}
\cite{Ferrario2007,Hurley2010}, namely, the
accretion induced collapse (AIC) channel.
AIC events occur when ultra-massive oxygen-neon (O-Ne) white dwarfs (WDs), accreting matter from their binary companions, approach the Chandrasekhar mass limit (${\sim} 1.4 M_\odot$), lose electron degeneracy pressure support due to electron capture on $^{24}$Mg and $^{24}$Na nuclei,
% \cite{Jones2019}, 
and collapse into neutron stars (NSs) 
%\cite{Miyaji1980, Takahashi2013, Schwab2015}
\cite{Miyaji1980}.
Conservation of angular momentum in AIC yields spin periods for the nascent NSs (with radii three orders of magnitude smaller than the parent WDs) of just a ${\sim}$few ms. 
Similarly, magnetic flux conservation during AIC means that progenitor
O-Ne WD magnetic fields $\sim 10^3 - 10^5 \ \rm G$
result in  field strengths  
of ${\sim} 10^{7}-10^{9}$ G for the newly born NSs, directly matching observations.
%without invoking hypothetical mechanisms for magnetic field decay. 
%
Furthermore, because little mass is lost in the collapse, the binary is not disrupted, nor does the system receive a significant natal kick \cite{Fryer1999,Kitaura2006,Tauris2013}. 
This means that NSs born via AIC have much smaller space velocities than those imparted to NSs during the generally asymmetrical explosion of a CCSN 
%\cite{Lyne1994, Wongwathanarat2013, Bear2018}
\cite{Lyne1994}.
Thus, MSPs born via AIC are likely to remain trapped in the bulge gravitational potential, allowing for a large population to build up over its history
and proffering an explanation of the detailed match %\cite{Macias2018,Bartels2018,Macias2019,Calore2021}
\cite{Macias2018,Bartels2018,Macias2019}
between  bulge stellar and GCE morphologies,
a match that is difficult to explain in the recycling channel in light of the large kicks typically delivered to core-collapse NSs (e.g., \cite{Ploeg2021}).

In this paper we present a binary population synthesis (BPS) forward model for the formation and evolution of the Galactic bulge AIC MSP population to ascertain whether its $\gamma$-ray phenomenology can explain the GCE (see Methods).
In this context, we note that template analyses \cite{Macias2018,Bartels2018,Macias2019} 
have uncovered distinct contributions to the
GCE signal from both 
the wider-scale ``Boxy'' bulge (BB) \cite{Freudenreich1998} and the  ``Nuclear Bulge'' (NB), 
the latter constituting
a stellar population
with a distinct star formation history concentrated on physical scales $\lesssim 200$\,pc around the Galactic centre. 
Here in the main text we concentrate on the former component which makes up around 90\% of the overall Galactic bulge mass.
We show in the Supplementary Information [S.I.]
that our model also successfully reproduces the measured
$\gamma$-ray emission from the NB
(see S.I.~sec.14)
.

%\section{AIC Evolution scenario} 
%\label{sec:Modeling}

O-Ne WDs are generated when stars develop carbon-oxygen (CO) cores in the range of ${\sim} 1.6-2.25\ \rm M_\odot$ that ignite carbon under semi-degenerate conditions.
Crucial to the formation of compact binaries \cite{Pac1976} is common envelope (CE) evolution which occurs when a star (generally a giant) overfills its Roche lobe engulfing both stars. 
A giant star with a convective envelope tends to further expand in response to mass loss triggering unstable mass transfer at rates so high the companion cannot capture all transferred matter.
Instead, matter accumulates in a common envelope surrounding both giant core and companion star \cite{deKool1992}. 
Since the envelope rotates more slowly than the stellar orbit, friction causes the stars to spiral together and transfer orbital energy to the envelope. 
This may release sufficient energy to shed the entire envelope, leaving either a close binary containing a WD and companion, or a coalescence of the giant's core with the companion (see S.I.~sec.1). 

In an AIC, while the rotational angular momentum of the collapsing O-Ne WD is largely retained, there is some angular momentum loss to gravitational wave (GW) radiation induced by the rapidly changing quadrupole moment.
Despite the negligible kick, some baryonic mass is suddenly lost,  lowering the effective gravitational mass which expands the binary. Depending on system parameters, the stars may come into contact again and mass transfer resume.
We find that donor stars can transfer more than $0.1 M_\odot$ of material to the NS and thus post-AIC accretion is a very important determinant of the spin evolution of an AIC MSP population.
%
%Given the $\gamma$-ray luminosity of a NS is powered by its rotational kinetic energy, 
Our BPS modelling includes all accretion torques acting on the NS  {see S.I.~sec.~4}.

%\section{Reference data set}
%\label{sec:Results}
%

%\subsection{Predicted Bulge MSP $\gamma$-ray emission}

Empirically, ${\sim} 10\%$ of an MSP's rotational kinetic energy  is lost to dipole radiation \cite{Abdo2013} which emerges at $\gamma$-ray wavelengths due to ``prompt'' emission by $e^+-e^-$ pairs within the magnetosphere.
A further, poorly-determined  fraction (1 -- 90\%: \cite{Sudoh2020,Song2021}), is carried away in relativistic pairs that escape the magnetosphere and, after possible re-acceleration at pulsar wind termination or intra-binary shocks,  generates additional synchrotron and $\gamma$-ray inverse Compton (IC) signals
on interstellar medium (ISM) fields.
Ultimately, both prompt and delayed signals are powered by the liberation, via magnetic dipole braking,
of rotational kinetic energy.
Fig.~1 shows the evolution of this spin down power according to our BPS model of the bulge AIC MSP population -- which thus sets an upper limit on the total (i.e., prompt + IC) $\gamma$-ray luminosity of the model MSP population -- together with a datum indicating the measured \cite{Macias2018} GCE luminosity.
There is sufficient power available to explain the signal at 13.8 Gyr cosmological (current) time.

We next determine the signal at Earth of $\gamma$-rays from the model bulge MSP population as described in Methods.
Our
initial estimate of the combined 
prompt spectrum detected from the bulge MSP population appears as the dot-dash blue curve in Fig.~2.
From this figure it is evident that the predicted prompt spectrum at Earth is promisingly similar to that measured and, in particular, does a good job of reproducing the amplitude of the ${\sim} 2$ GeV bump.
However, 
i) the amplitude of the central value of the predicted signal is somewhat low and ii) the spectral shape is not a good fit to high energy ($E_\gamma \gtrsim 10$\,GeV) 
tail with an overall $\chi^2$ of 346 for 15 data points (and no tunable parameters), where we account for the errors in both the predicted prompt spectrum and the spectral measurements.
But there is a potential further source of $\gamma$-ray emission powered by the MSP spin down that could contribute, in particular, to this tail: IC off photons of the interstellar radiation field
by $e^{\pm}$ that escape MSP magnetospheres
(e.g.\ \cite{Yuan2015}).
%(e.g.\ \cite{Yuan2015,Petrovic2015,Calore2015,Horiuchi2016,Linden2016,Song2019,Song2021,Macias2021}). 
%
%This process can supply spectrally hard emission up to $\gtrsim 30$\,GeV, as required (e.g., \cite{Song2021}).  
%

Now, while the total prompt emission  is prescribed (up to statistical fluctuations) once we have the model %$\{P_i,\dot{P}_{\rm MB,i}\}$ 
MSPs' periods and period (temporal) derivatives
from our BPS, this is not the case for the delayed IC signal which is subject to a number of factors not determined within the model (see S.I.~sec.~7).
On the other hand, 
we can exploit two significant constraints: i)  synchrotron radiation off the bulge magnetic field by the population of CR $e^\pm$ escaping MSPs 
(and supplying the IC signal)
should not exceed the microwave frequency flux density measured from the inner Galaxy \cite{Ade2013} and ii) the overall prompt + IC + synchrotron luminosity may not exceed the total power liberated by magnetic braking. 

Using these constraints, we perform a simultaneous fitting of the IC and prompt signals to the GCE data points allowing the amplitude of the prompt signal to float within the  uncertainty range around the initial estimate; the fitted spectra and their uncertainty bands are shown in Fig.~2.
Evidently, an old population of MSPs can well reproduce  the GCE $\gamma$-ray signal.
%
%\autoref{fig:BB_spec} 
Fig.~3
shows the broad-band, non-thermal emission from our fit.
Here we saturate by construction the microwave data points \cite{Ade2013} that describe the spectrum of the ``microwave haze'', 
an enduringly mysterious, non-thermal emission feature of the inner Galaxy
%first discovered in WMAP data \cite{Finkbeiner2004} 
for which some have considered an origin in inner Galaxy pulsars or MSPs
%\cite{Zhang2009,Kaplinghat2009,Malyshev2010,Harding2010,McQuinn2011}.
\cite{Zhang2009}.
Our fit assumes that a significant fraction, close to 100\%, of the spin-down power liberated by magnetic braking ends up in relativistic pairs (cf.\cite{Sudoh2020}).

%\subsection{Population of $\gamma-$ray resolvable Bulge MSPs}

%\section{Discussion}
%\label{sec:Conclusion}

We have evolved a model population of Galactic bulge binaries to the present day; these generate  ${\sim} 1.1\times10^5$ accretion induced collapse (AIC) neutron stars that qualify as MSPs ($P< 40$ ms).
Individually, these MSPs are indistinguishable from those from ``recycling''.
% \citep[e.g.,][]{Tauris2013}.
%
At a population level, bulge MSPs are older and thus dimmer than typical local MSPs and 
characterised by a luminosity distribution
$10^{31\pm 3}$ erg/s
including both prompt and IC components (and taking the mean and standard deviation in the log).
The model bulge AIC MSP population can fully explain both the spectrum and overall luminosity of the GCE and, as a bonus, the microwave haze from the inner Galaxy.
Our model does not predict too many bright/resolvable  MSPs (S.I.~sec.~11);
nor too many bright, individual X-ray sources; nor an excessive integrated X-ray luminosity (S.I.~sec.~12). 

The upcoming Cherenkov Telescope Array (CTA) will look at the sky in photons of energies $20 \ \rm GeV$ to $300 \ \rm TeV$ and is expected to spend at least 500 hours observing the inner Galaxy \cite{CTA2019} in which case we expect its point source sensitivity at 0.1\,TeV to approach few $\times 10^{-13}$ erg cm$^{-2}$ s$^{-1}$ \cite{Maier2019}.
Thus, CTA should resolve at least a handful of GCE MSPs
according to the best-fit spectral parameters reported above. %(cf.~\cite{Macias2021}).
In addition,
we expect point source inference techniques directed at probing the flux distribution of the sub-threshold population 
in \fermi \ data
[e.g.,\cite{Buschmann2020}]
will,
as they become robust to systematics, 
start to test our MSP population predictions.

%\newpage

%\printbibliography[segment=\therefsegment,check=onlynew]
%\printbibliography[segment=\therefsegment]

%\bibliography{sgrBib}

\clearpage

\printbibliography[segment=\therefsegment,title={References}, check=onlynew]

\newpage

\section*{Figures}

%\subsection{Figure 1}
%\label{fig:L_gamma_comp}

\begin{figure}[ht]
\adjustimage{width=0.9\columnwidth}{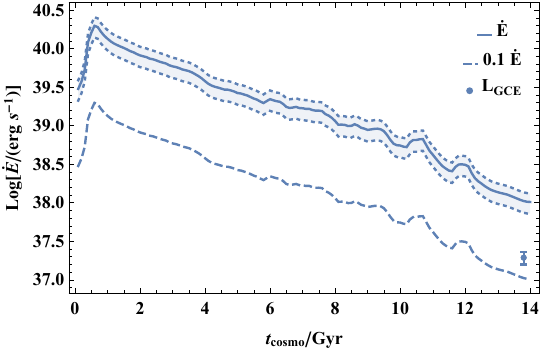}
            % \decoRule
            \caption{{\bf Spin-down power liberated by magnetic dipole braking of Galactic bulge millisecond pulsars over cosmological history according to our BPS model.}
Curves are as follows:
{\bf solid:}
total spin-down power (the band indicates $\pm1\sigma$ error, dominated by the uncertainties in the bulge binary fraction and the bulge stellar mass);
{\bf dashed:}
(an indicative)  10\% of the spin-down power.
The time series have been smoothed over 0.5 Gyr.
{\bf Data point:} The luminosity datum at 13.8 Gyr, $2.2 \pm 0.4 \times 10^{37}$ erg/s ($\pm 1\sigma$), is inferred from the measured \cite{Macias2019} Boxy Bulge spectrum  together with a weighted mean distance (as obtained from Monte Carlo sampling of the spatial model \cite{Ploeg2020}) to each MSP of $\langle d_{\rm MSP} \rangle = 7.89$  kpc).}\label{fig:L_gamma_comp}
\end{figure}

\begin{figure}[ht]
            \adjustimage{width=1\columnwidth}{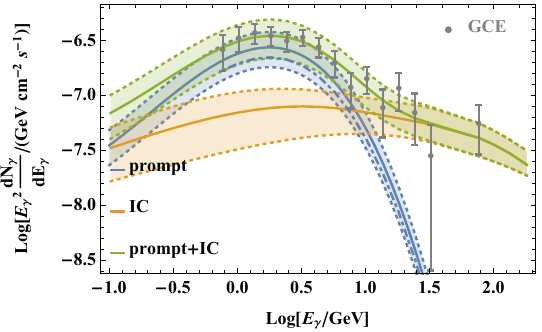}
            % \decoRule
            \caption{{\bf GCE spectrum at the Earth.}
Data points display the measured GCE spectrum
from ref.~\cite{Macias2019} (error bars indicate $\pm 1\sigma$).
The curves show the fitted $\gamma$-ray spectrum at Earth of the model MSP population at 13.8\,Gyr and are as labelled in the figure.
The {\bf dot-dash blue line} shows the central value directly predicted by our BPS.
All solid lines have been fitted following the procedure outlined in S.I.~sec.~6 (bands indicate $\pm1\sigma$ error,  determined from application of the profile likelihood method as presented in the S.~I.~sec.~8).
The spectral index of the CR $e^\pm$ population is $\gamma_e = -2.1^{+0.1}_{-0.2}$ and its cutoff energy is $(9^{+10}_{-2}) \times 10^{11}$\;eV.}\label{fig:L_spec}
\end{figure}

\newpage

\begin{figure}[ht]
            \adjustimage{width=1\columnwidth}{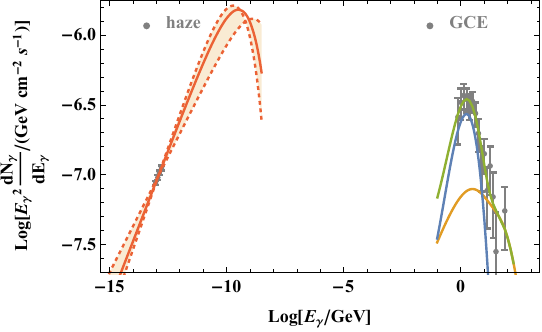}
            % \decoRule
            \caption{{\bf Broad-band, non-thermal spectrum of the Galactic bulge region.}
{\bf Grey data points}
show (left, `haze') the measured spectrum of the
microwave haze \cite{Ade2013}
and (right, `GCE') the GCE \cite{Macias2019}
(error bars indicate $\pm 1\sigma$).
The fitted curves are as follows:
{\bf green:} total $\gamma$-ray emission;
{\bf blue:} prompt $\gamma$-ray emission;
{\bf yellow:} IC $\gamma$-ray emission;
and
{\bf orange:} synchrotron emission.
These curves are for spectra at Earth of the model MSP population at 13.8\,Gyr. 
The colour band indicates the $\pm1\sigma$ error 
(determined from application of the profile likelihood method as presented in the S.~I.~sec.~8)
on the fitted synchrotron spectrum (error bands on the $\gamma$-ray curves are suppressed for clarity, but are as shown in Fig.~2); all solid lines have been fitted following S.I.~sec.~6.
Best fit parameters are as given in the caption of Fig.~2 plus a magnetic field for the bulge $(2.3^{+ 0.6}_{ - 0.4}) \times 10^{-5}\ $G.}\label{fig:BB_spec}
\end{figure}

\clearpage

\newrefsegment

\section*{Methods}
\label{sec:methods}

\section*{Binary-star evolution (BSE) code}

To evolve a model population of binaries up to AIC, we have used the  binary-star evolution (BSE) code 
%\cite{Hurley2000, Hurley2002, Hurley2013}
\cite{Hurley2000}. 
The BSE algorithm presented by ref.~\cite{Hurley2002}, with updates as described by ref.~\cite{kiel_hurley_2006}, is a state-of-the-art rapid evolution algorithm that melds single star evolution \cite{Hurley2000, Hurley2002} with binary evolution. 
BSE takes as input stellar zero age main sequence (ZAMS) masses, $M_{\rm 1,ZAMS}$ and $M_{\rm 2,ZAMS}$ for the primary and secondary respectively ($M_{\rm 1,ZAMS}$ $\geq$ $M_{\rm 2,ZAMS}$), and the initial binary separation, $a$.

BSE evolves binary stellar components through detached, semidetached, or contact phases of binary evolution. 
During the evolution of a binary system, eccentricity may evolve, and, owing to tidal interactions (which are modeled in detail within BSE), synchronization of stellar rotation with the orbital motion may occur. 
BSE includes all orbital angular momentum loss mechanisms including gravitational radiation and mass loss. 
Wind accretion (wherein the secondary accretes some of the material lost from the primary in a wind), and Roche lobe overflow (where mass transfer occurs if either star fills its Roche lobe), are modelled on nuclear, thermal, or dynamical time-scales -- whichever is smallest.
BSE dynamically adjusts orbital parameters while taking into consideration any mass variation in the two stars. 
Stars with deep surface convection zones and degenerate stars are unstable to dynamical time-scale mass loss unless the mass ratio of the system is less than some critical value. The outcome is a common-envelope event \cite{Pac1976}. 
This results in either the merging of the two stars, likely accompanied by the generation of a very strong magnetic field  \cite{Wick2014, Briggs2015}, or the formation of a close binary. 
Conversely, mass transfer on a nuclear or thermal time-scale is assumed to be a steady process. 
Prescriptions to determine the type and rate of mass transfer, the response of the secondary to accretion, and the outcome of any merger events are in place in BSE; details can be found in refs.~\cite{Hurley2000, Hurley2002, Hurley2013}.

\subsection*{Model binary population}

Our synthetic population consists of ${\sim} 6.3\times 10^7$ binary systems which, given parameter choices, would be hosted by a stellar  population of $2.0 \times 10^9 M_\odot$ total  ZAMS mass.
The binaries are numerically evolved up to 14 Gyrs generating a total of ${\sim} 9.1\times 10^3$ AIC events, which start when the population is only ${\sim} 0.2$ Gyr old; c.f.~Extended Data (E.D.) Fig.~1. 
In order to construct our model binary population we adopt empirically-motivated, ``off-the-shelf'' 
parameter choices (see next sub-section).

\subsection*{Choice of initial binary parameters}

Before a detailed description of each characteristic or parameter choice we make, we provide a brief summary of our model binary population:
i) A binary star fraction of 70\% is adopted \cite{Ruiter2019};
ii) $M_{1,ZAMS}$'s are drawn from the initial mass function of \cite{Kroupa1993};
iii) $M_{\rm 2,ZAMS}$'s are drawn from a flat mass ratio distribution \cite{Ferrario2012,Hurley2002};
iv) Initial eccentricities are set to zero;
v) Orbital periods at zero age are drawn from a log-uniform distribution covering $10$ to $10^4$ days \cite{DucheneKraus2013};
vi) Common envelope efficiency factor (see S.I.~sec.~1) is set to $\alpha_{\rm CE}=1$ \cite{Ruiter2019};
vii) The binding energy parameter $\lambda_{\rm b}$
is calculated using the Cambridge STARS code \cite{Eggleton1971, Pols1995, Pols98};
and viii) Magnetic fields are assigned to each nascent NS via random sampling from the log-normal distribution of \cite{Ploeg2020}.

{\bf Secondary masses:} It is standard practice in numerical binary population synthesis calculations and supported by observations of binary systems in the Galaxy \cite{Moe2013,Mazeh2003}, that companion secondary masses be described by a flat distribution relative to primary masses: $0<q=M_{2,ZAMS}/M_{1,ZAMS}<1$.

{\bf Initial eccentricity:} 
Because the detailed calculations of \cite{Hurley2002} have shown that orbits circularise before Roche lobe overflow occurs and that it is not necessary to include a distribution of eccentricities in population synthesis of interacting binaries, in our BPS we initialise all binaries with circular orbits (see also \cite{Willems2002,Hurley2010, Moe2017,Ruiter2019}).

\subsection*{A typical evolutionary pathway towards AIC}
    
(The reader is referred to E.D.~Fig.~2).
The primary's ZAMS mass is $7.636\ \rm M_\odot$ and its radius at $t=0$ is $3.364\ R_\odot$. The secondary has a ZAMS mass of $5.702\ \rm M_\odot$ and a radius of $2.841\ R_\odot$. The initial separation at the time of the system's birth is $2369.2 \ R_\odot$. The primary evolves through HG, First Giant Branch, CHeB, EAGB and HeGB phases before becoming a O-Ne WD. 
    
The first interaction occurs at ${\sim} 44$ Myr after starburst. Initially, the ZAMS primary undergoes Roche Lobe Overflow (RLOF) as it expands on the main sequence and the main sequence secondary accretes matter stably. At ${\sim} 54$ Myrs from birth, the primary evolves through the Hertzsprung gap and then along the asymptotic giant branch, all the while donating mass to the MS secondary. This is followed by a common envelope (CE) phase between the late AGB primary ($6.859\ \rm M_\odot$) and the main sequence companion ($5.434\ \rm M_\odot$). The CE phase causes the binary orbit to shrink and the primary moves from the EAGB to become a Giant Branch Naked Helium star and soon evolves into a O-Ne WD. 
        
The next interaction does not occur until there is stable RLOF (which begins at time $t {\sim} 83$ Myrs) between the Hertzsprung Gap secondary ($5.969\ \rm M_\odot$) and the O-Ne WD ($1.315\ \rm M_\odot$). The primary continues accreting mass for another ${\sim} 20$ Myrs ($t = 103$ Myrs) until the secondary (now a Hertzsprung Gap Naked Helium star) collapses into a CO WD ($0.683\ \rm M_\odot$). The system remains dormant till $t=108.4$ Myrs, while spiraling closer ($a= 0.194 R_\odot$) via gravitational radiation. The dormant phase is followed by the O-Ne WD accreting mass from the CO WD for ${\sim} 0.3$ Myrs (an AM Can\,Ven system but with a much more massive WD primary \cite{Provencal1997}). Once the O-Ne WD ($1.439 \rm M_\odot$,$5.3\times 10^3$ s) approaches the Chandrasekhar limit for rotationally supported WDs, it undergoes AIC to form a NS ($1.357\ \rm M_\odot$, $P=4.8$ ms) at $t = 109$ Myrs from birth. After it is formed, magnetic dipole braking is the dominant spin-down mechanism for the NS for ${\sim} 10^5$ yrs, and soon after that, the NS begins accreting mass from the CO WD ($0.109\ \rm M_\odot$).

E.D.~Fig.~3 shows the number of systems that undergo an AIC event in our simulated population and their corresponding companion stellar types. Each line represents an AIC event. Regions of solid colour are regions of high number density of systems with respective donor types that undergo an AIC event at any point in the age of the Bulge. Systems with CO WD, He WD and He-MS donors dominate the AIC progenitor population. We also observe that a population of NSs formed with Core Helium Burning (CHeB) stars and WDs as companions have the longest delay times which extend even beyond a Hubble time; RG and MS donors show similar trends.

\subsection*{Merger Induced Collapse}
%\label{sec:MIC}

For completeness, we mention that in the present study we have neglected Merger-Induced Collapse (MIC) events that occur when two massive WDs merge due to gravitational wave radiation \cite{Ivanova2008}. 
This is because they occur relatively early post star-formation \cite{Ruiter2019,Lyutikov2019}, supply less than 10\% of the MSP population \cite{Ruiter2019}, and, in any case, produce nascent NSs with super-strong magnetic fields \cite{Wick2014} that quickly spin down and no longer qualify as MSPs today.

\subsection*{Application to Galactic bulge}
%\label{sec:Applic}

In order to apply our model results to the Galactic bulge, we need to scale up from the 
$2.0 \times 10^9 \ M_\odot$ ZAMS equivalent, host stellar mass of the BPS to the bulge
ZAMS mass.
The latter quantity can be
inferred from its {\it current} measured \cite{Portail2015} stellar mass $(1.55 \pm 0.15) \times 10^{10} \rm M_\odot$ (including stellar remnants), its star formation history,
and a model for stellar mass loss in winds and ejecta
\cite{Maraston1998}.
The current bulge stellar mass
needs rescaling upwards by $1.81 \pm 0.02$
to obtain the ZAMS mass (where this error range purely reflects the uncertainty arising from the star formation history), leading to an overall rescaling from 
the ZAMS mass implied by the modeled binary population to the
bulge of
$14.0 \pm 4.1$. 
This implies  $(1.3 \pm 0.4) \times 10^5$ AIC events have occurred in the bulge up to now (cf.~E.D.~Fig.~1) 
with 
$(1.1 \pm 0.3) \times 10^5$ MSPs currently in the bulge, and an AIC expected every few Myr.

\subsection*{Calculation of model MSP $\gamma$-ray spectra}

Given that our BPS gives us the instantaneous period and period derivative of each MSP, we use an empirical prompt $\gamma$-ray luminosity model of the form  \cite{Gonthier2018}:
\begin{equation}
    L_{\rm \gamma, prompt,i} = 
    f_{\rm \gamma, prompt} 
\left(P_i/{\rm s}\right)^{\alpha_\gamma}
    \dot{P}_{\rm MB,i}^{\beta_\gamma} 
    \ \rm erg \  \ s^{-1},
\label{eq:Lgamma}
\end{equation}
where $P_i$ is the period of the $i$th MSP 
and $\dot{P}_{\rm MB,i}$
is the time derivative of its period due to magnetic dipole braking.
We distribute this over a spectrum  \cite{Ploeg2020}:
\begin{equation}
    \frac{dN_{\rm \gamma,prompt,i}}{dE_\gamma} \propto E_\gamma^{-\Gamma_i} e^{(-E_\gamma/E_{\rm cut,i})^{2/3}} \, .
\label{eq:Lgamma_spectrumInit}
\end{equation}
Then the prompt flux received from each model MSP
is determined from
\begin{equation}
     F_{\rm \gamma,prompt,i} \equiv
    \frac{L_{\rm \gamma,prompt,i}}{4 \pi d_i^2} \,.
\label{eq:KnormInit}
\end{equation}
Finally, after assigning the required distances
and the spectral and other parameters (S.I.~sec.~6), 
we can generate an {\it initial} estimate of
the combined prompt emission from the bulge MSP population.

\clearpage

\subsection*{Data availability}

The model MSP data set created with our code  has been posted to Zenodo  at DOI 10.5281/zenodo.6342560.

\subsection*{Code availability}

Our code is based on the BSE\cite{Hurley2010} code available here:  \url{http://astronomy.swin.edu.au/~jhurley/bsedload.html}.
We have made some refinements and modifications to BSE to account for the spin evolution of AIC-formed neutron stars post formation. These code extensions are presented in a GitHub repository here: \url{https://github.com/gautam-404/Binary-Evolution/tree/master}.

%\printbibliography[segment=\therefsegment, check=onlynew]

\section*{Acknowledgements}

RMC thanks Mark Krumholz, Ivo Seitenzahl, Gavin Rowell, Celine Boehm, 
Matthew Baring, and Ciaran O'Hare for enlightening conversations.
Nick Rodd and Shunsaku Horiuchi are thanked for a close reading of the  paper in draft, and Tracy Slatyer for comments.
RMC acknowledges 
support from the Australian Government through the Australian Research Council, award
DP190101258 (shared with Prof. Mark Krumholz).
AJR has been supported by the Australian Research Council through grant number FT170100243.  
Parts of this research were undertaken with the assistance of resources
and services from the National Computational Infrastructure (NCI), which is supported
by the Australian Government, through the National Computational Merit Allocation
Scheme and the UNSW HPC Resource Allocation Scheme. The work of O.M. was supported by JSPS KAKENHI Grant Number JP20K14463 and by the World Premier International Research Center Initiative (WPI Initiative), MEXT, Japan.

\section*{Author contributions}

RMC conceived the project in consultation with LF and AJR. AG performed the BPS under the supervision of LF, AJR, and  RMC.
AG added new functionality to the existing BSE code to model neutron star period evolution under accretion torques. RMC and HP analysed the BPS data to derive model $\gamma$-ray luminosities and spectra. CG consulted about data and statistical analysis. OM performed a number of novel $\gamma$-ray template analyses of the GCE in support of the project. An original draft of the paper was written by AG. This was subsequently amended and extended by RMC in consultation with all the other authors.

%\section*{Additional information}

%To include, in this order: \textbf{Accession codes} (where applicable); . 

\section*{Competing interests}

The authors declare no competing interests.

\newpage

\printbibliography[segment=\therefsegment,title={Methods References}, check=onlynew]

\clearpage

\newrefsegment

\section*{Extended Data}
\setcounter{figure}{0}
\setcounter{table}{0}

\renewcommand{\figurename}{Extended Data Figure}

\begin{figure}[ht]
            \adjustimage{width=1\columnwidth}{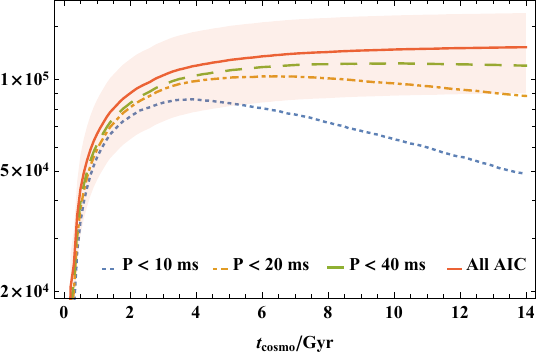}
            % \decoRule
            \caption{{\bf Cumulative AIC events and number of model bulge MSPs
        over cosmological time}.
        MSP periods $P$ are as labelled in the legend; we define any NS with $P < 40$\,ms  as an MSP. 
        The $\pm1\sigma$ error band on the red curve for all AIC events reflects the  uncertainties stemming from the bulge stellar mass determination and binarity fraction. (For clarity, equivalent error bands on the other curves are not shown.)}\label{fig:ED1}
\end{figure}

\begin{figure}[ht]
            \adjustimage{width=0.8\columnwidth}{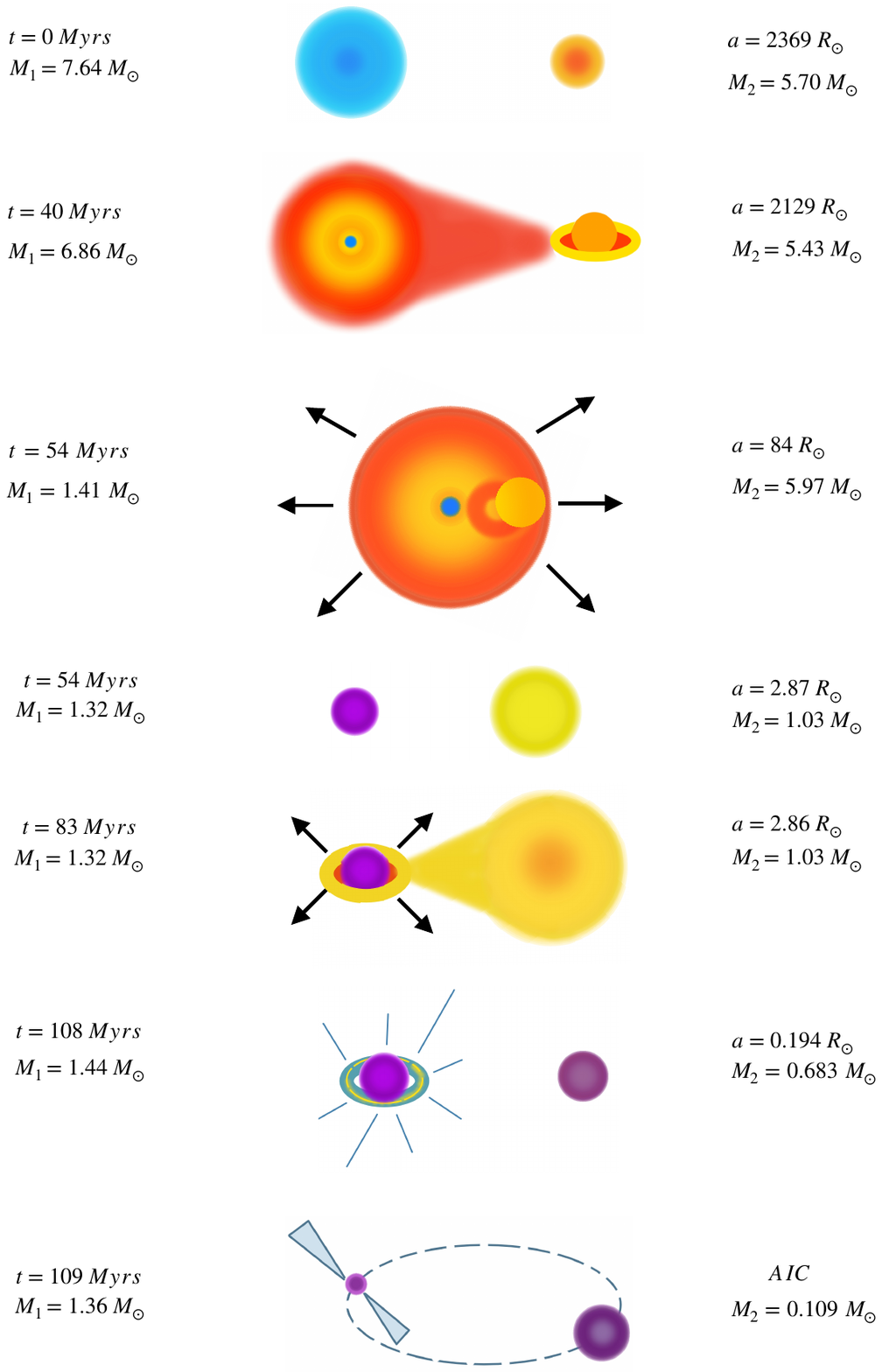}
            % \decoRule
            \caption{{\bf The main evolutionary stages towards, and beyond, accretion induced collapse of a white dwarf.} This schematic is for the model binary whose history is described in the Methods section `A typical evolutionary pathway towards AIC'.}\label{fig:ED2}
\end{figure}

\begin{figure}[ht]
            \adjustimage{width=1\columnwidth}{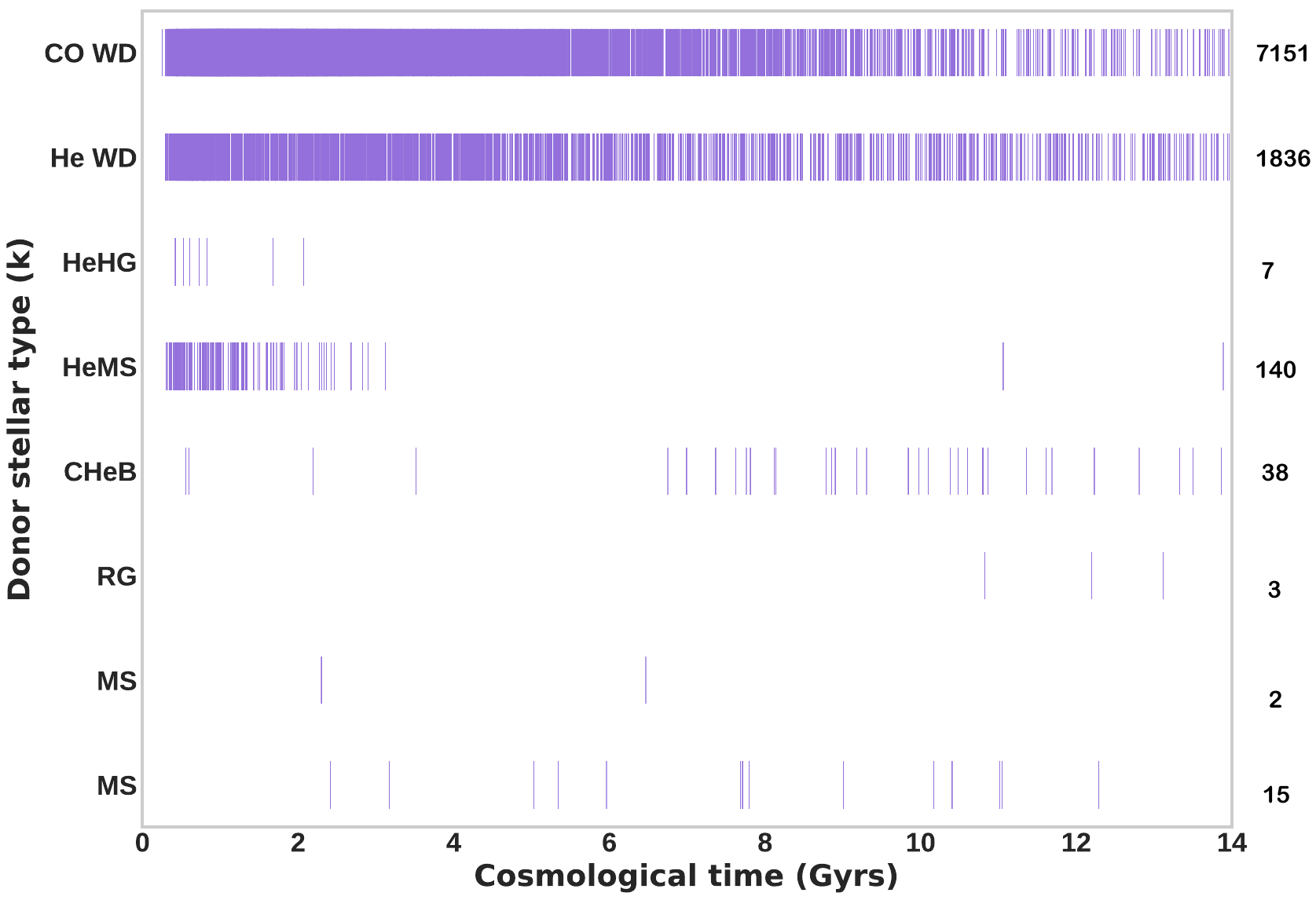}
            % \decoRule
            \caption{{\bf  Cosmological time vs. donor star type at time of AIC for all AIC events in our simulated population.}
        Given our empirically-motivated parameter choices, this model binary population is as expected for a host stellar population of total zero age main sequence mass
        of $2\times10^9 \ M_\odot$. For more details on stellar types, the reader is referred to S.I.~sec.~2.}\label{fig:ED3}
\end{figure}

\clearpage

\setcounter{figure}{0}
\setcounter{table}{0}

\renewcommand{\figurename}{Supplementary Information Figure}

\section*{Supplementary Information for {\it Millisecond Pulsars from Accretion Induced
Collapse as the Origin of the Galactic Centre
Gamma-ray Excess Signal}, Gautam et al. (2022)}

\section{Common envelope treatment}
\label{sec:CE}

BSE's CE evolution follows closely the description of ref.~\cite{Tout1997} where the outcome depends on the initial binding energy, $E_{\rm bind}$, of the envelope and the initial orbital energy, $E_{\rm orb}$, of the two cores. 
A structure parameter $\lambda$  (also called the envelope binding energy factor) is introduced so that the binding energy of the envelope of a star of mass $M$, degenerate core mass $M_c$ and radius $R$ is \cite{Webbink1984}
\[
E_{\rm bind}=\frac{G M (M - M_c)}{\lambda R},
\]
where $G$ is the gravitational constant. Because the ejection of the envelope may not be complete, an efficiency factor, $\alpha_{CE}$, is introduced. Thus
\[
\Delta E_{\rm orb}=\alpha_{CE}E_{\rm bind}
\]
gives the fraction of the orbital energy that is transferred to the envelope.

Most CE evolution models use $\alpha_{CE} = 1$ \cite{Willems2002, Pfahl2003} and this is the value that we also use in our modeling. However, there is a lot of uncertainty in this value \cite[e.g.][]{Tutukov1996} due to the CE evolutionary phase being poorly understood, mostly because it happens on dynamical time scales and therefore there are no observations that can be used to constrain the physical parameters that dominate the CE evolution. The parameter $\lambda$ is of the order of unity and depends on the stellar density distribution. We have used the BSE prescription for $\lambda$ calculated with the Cambridge STARS code \cite{Eggleton1971, Pols1995} and assumed that half of the envelope's internal energy aids its ejection. This setup is very similar to ``Model 2'' of ref.~\cite{Ruiter2019} and it
results in the product $\alpha_{CE} \lambda$ varying by  $\lesssim$ one order of magnitude depending on binary parameters.
Note that in their ``Model 1'', ref.~\cite{Ruiter2019} also explored a ``classical'' CE prescription for which the product is held constant: $\alpha_{CE} \lambda = 1$. 
Overall, while these different treatments of the CE phase led to differences in the relative importance of various, individual channels to the production of AIC events, the overall AIC rate was similar for Models 1 and 2 (to within $\sim$30\%; see section 3.2 of ref.~\cite{Ruiter2019}).

\section{Stellar types}
\label{sec:stellar_types}
    
We use the stellar types ($k$) correspond to the evolutionary phase designated by ref.~\cite{Hurley2000}, which are:
    \begin{align*}
        0 &\rightarrow \text{ Main Sequence (MS)}\ M \le 0.7 \rm M_\odot \\
        1 &\rightarrow \text{ MS star}\ M >0.7\rm M_\odot\\
        2 &\rightarrow \text{ Hertzsprung gap (HG)}\\
        3 &\rightarrow \text{ Red Giant Branch (RG)} \\
        4 &\rightarrow \text{ Core Helium Burning (CHeB)} \\
        5 &\rightarrow \text{ Early Asymptotic Giant Branch (EAGB)} \\
        6 &\rightarrow \text{ Thermally Pulsing AGB (TPAGB)} \\
        7 &\rightarrow \text{ Helium Main Sequence (HeMS)} \\
        8 &\rightarrow \text{ Helium Hertzsprung gap (HeHG)} \\
        9 &\rightarrow \text{ Helium Giant Branch (HeGB)} \\
        10 &\rightarrow \text{ Helium White Dwarf (He WD)} \\
        11 &\rightarrow \text{ Carbon/Oxygen White Dwarf (CO WD)} \\
        12 &\rightarrow \text{ Oxygen/Neon White Dwarf (O-Ne WD)} \\
        13 &\rightarrow \text{ Neutron Star (NS)} \\
        14 &\rightarrow \text{ Black Hole (BH)} \\
        15 &\rightarrow \text{ Massless remnant (after SN Ia explosion)}
    \end{align*}

\section{Angular momentum evolution in AIC and gravitational radiation}
\label{sec:L}

The period of the new-born AIC NS due to conservation of angular momentum of the collapsing WD is given by,
\begin{equation}
    P_{NS}= \bigg(\frac{R_{NS}}{R_{WD}}\bigg)^2 \omega_{WD}.
\end{equation}

The angular momentum loss due to gravitational radiation is accounted for following ref.~\cite{Watts_Andersson2002}.
The total stellar angular momentum $J$ is the sum of the star's angular momentum $J$ and the canonical angular momentum of the r-mode $J_c$
\cite{Watts_Andersson2002}: 
\begin{equation}
    J_c = -\frac{3 \Omega \alpha^2 \tilde{J} M R^2}{2}
\end{equation}
where, for an $n = 1$ polytrope, $\tilde{J} = 1.635 \times 10^{-2}$ and  $\alpha$ is the mode amplitude \cite{LindblomOwen_1998}.
The total torque on the star $\dot{J}$ can therefore be written as
\begin{equation}
    \dot{J} = \dot{I} \Omega + I\dot{\Omega} + \dot{J_c}.
\end{equation}
Here, $\Omega$ is the star’s angular velocity. $I = \tilde{I}MR^2$ is the moment of inertia of the star. 
Because the star is modelled as an $n = 1$ polytrope, then $ \tilde{I} = 0.261$. 
For this work, we only account for radiation from the dominant $l = m = 2$ current multipole of the r-mode. 
Following \cite{Watts_Andersson2002}, we assume that the canonical angular momentum evolves as 
\begin{equation}
    \dot{J_c} = -\frac{2 J_c}{\tau}
\end{equation}
where 
\begin{equation}
    \tau^{-1} = -|\tau_g|^{-1} + \tau_b^{-1} + \tau_s^{-1}
\end{equation}
and $\tau_g$ is the timescales associated with growth of the linear perturbation due to gravitational radiation back reaction and $\tau_{b,s}$ are the timescales required for the dissipation of energy due to bulk and shear viscosity, respectively. The timescales due to the $l = m = 2$ current multipole are:
\begin{equation}
    \tau_g \approx -47 M_{1.4}^{-1} R_{10}^{-4} P_{-3}^6 \ \rm{s}
\end{equation}
\begin{equation}
    \tau_b \approx 2.7\times10^{11} M_{1.4} R_{10}^{-1} P_{-3}^2 T_{9}^{-6} \ \rm{s}
\end{equation}
\begin{equation}
    \tau_s = 6.7 \times 10^{7} M_{1.4}^{-5/4} R_{10}^{23/4} T_9^{2} \ \rm{s}
\end{equation}
where $M_{1.4} = M/1.4 \ \rm{M_\odot}$, $R_{10} = R/10 \ \rm{km}$, $P_{-3} = P/1 \ \rm{ms}$ and $T_9 = T/10^9 \ \rm{K}$.

The evolution of the mode amplitude, $\dot{\alpha}$, and the angular velocity, $\dot{\Omega}$, are expressed as
\begin{equation}
    \dot{\alpha} = -\alpha \bigg(\frac{1}{\tau} + \frac{\dot{\Omega}}{2\Omega} + \frac{\dot{M}}{2M} \bigg)\qquad\mbox{and}\qquad   \dot{\Omega} = \bigg(\frac{\dot{J}}{I} - \frac{\dot{M}\Omega}{M} + \frac{3 \tilde{J} \alpha^2 \Omega}{\tilde{I} \tau} \bigg)
\end{equation}

\section{Spin evolution of neutron stars post AIC}
\label{sec:spin}
In this section we outline our modelling procedure to compute the spin evolution of post-AIC NSs in the non-accretion and accretion phases. 

\subsection{Non accreting neutron stars} 
If there is no accretion onto the NS, this spins down via magnetic dipole radiation \cite{possenti2000},
\begin{equation}
        \dot{P}_{\rm MB} \simeq 3.1 \times 10^{-8}\mu_{30}^2I_{45}^{-1}\bigg(\frac{P}{\rm s}\bigg)^{-1}\,\, \frac{\rm s}{\rm yr} \, ,
\label{eq:PdotEM}
\end{equation}
where $P$ is the NS rotation period, $\mu_{30}$ is its magnetic moment ($\mu = B R_{NS}^3$) in units of $10^{30}\ \rm G\ cm^3$ and $I_{45}$ is the NS's moment of inertia ($I \simeq 2/5 M_{NS} R_{NS}^2$) in units of $10^{45}$\,g\,cm$^2$.
 We take into account the effects of dipole braking only on the spin evolution of non-accreting pulsars, whereas, in the case of accreting NSs in binaries, we also include the torques applied by the in-falling matter on the NS.
\subsection{Accreting neutron stars}

Accreting NSs are called low mass X-ray binaries (LMXBs). In these systems the companion star overfills its Roche lobe and matter escapes through the inner Lagrange point and flows toward the NS's magnetosphere. 
Because this matter has a large specific angular momentum, it cannot impact directly on to the surface of the NS but instead forms an accretion disk. 
Over time, matter in the accretion disk drifts gradually inwards as angular momentum is transferred outwards by viscous torques. 
Once the material reaches the magnetospheric radius $r_m$  where the magnetic pressure dominates over the gas ram pressure \cite{Gittins2019},
\begin{equation}
     r_m = \xi \bigg( \frac{\mu^4}{2GM\dot{M}^2} \bigg)^{1/7},
\label{eq:r_m}
\end{equation}
the gas is magnetically channelled along magnetic fields lines to impact near the magnetic poles of the NS. %
Here $0.5\le\xi\le 1.4$ depends on the fraction of the star's magnetic flux threading the disk.  \cite{Gittins2019}; 
in our work we adopt $\xi = 0.5$ \cite{Wang96_magthreading}.

At the magnetospheric boundary, the coupling between the field lines and the disk results in a torque acting on the NS that can spin it up or down depending on whether $r_m$ is larger or smaller than the co-rotation radius $r_c$ given by
\begin{equation}
   r_c = \bigg(\frac{GM}{\Omega^2} \bigg)^{1/3}.
\end{equation}
At $r_c$ matter rotates with an angular velocity equal to the spin angular velocity $\Omega= 2\pi/P$  of the NS.

Thus, 
there are two possible cases that need to be considered \cite{Lipunov1992, Boldin2010, Gittins2019,GhoshLamb1979ii, GhoshLamb1979iii}, namely:

\begin{enumerate}
    \item If $r_m<r_c$, the standard torque (neglecting magnetic field effects, such as distortion and/or pinching of the magnetic field lines at the threading region) is  \cite{GhoshLamb1979ii} 
    \begin{equation}
        \dot{\Omega}= \frac{\dot{M}\ r_m^2 \ \Omega_K(r_m)}{I} = \frac{\dot{M} \sqrt{GMr_m}}{I}
        \, ,
    \end{equation}
    where $\Omega_K(r_m)$ is the Keplerian angular velocity at $r_m$ and $I$ is the moment of inertia of the NS. The disk-star system is thus regarded as an accretor. Here the gas channelled on to the NS has greater specific angular momentum than the NS and thus the star is spun up.
    
    \item When the accretion disk is truncated at a radius larger than the co-rotation radius ($r_m>r_c$), the disk-star system is regarded as a ``propeller''. In this phase, the sign of $(1-\omega_s)$ 
    dictates whether the NS accretes the gas and spins up (the ``slow rotator'' regime, $\omega_s < 1$) or ejects the gas and spins down (the ``fast rotator'' regime, $\omega_s > 1$ \cite{Wang96_magthreading}). 
    The quantity $\omega_s$ is the fastness parameter, defined as the ratio of the spin frequency of the NS
    ($\Omega$) to the Keplerian orbital frequency ($\Omega_K(r_m)$) at the magnetospheric boundary:
    \begin{equation}
        \omega_s = \frac{\Omega}{\Omega_K(r_m)} \, .
    \end{equation}

    The change in spin period is given by,
    \begin{eqnarray}
       \dot{P} &\approx& -8.1 \times 10^{-5}\ \xi^{1/2} \ M_{1.4}^{3/7} \ I_{45}^{-1} \ \mu_{30}^{2/7} \bigg[ \bigg( \frac{P}{1s} \bigg) \ \dot{M}_{-9}^{3/7} \bigg]^2 \ (1-\omega_s)\,\,\rm s \ yr^{-1}\, ,\nonumber 
    \end{eqnarray}
    where $\dot{M}_{-9} = \dot{M}/10^{-9}$\,M$_\odot$\,yr$^{-1}$
\end{enumerate}
    
\subsection{Spin period evolution of bulge MSP population}
    
\begin{figure}[ht]
        \centering
        \adjustimage{width=0.8\linewidth}{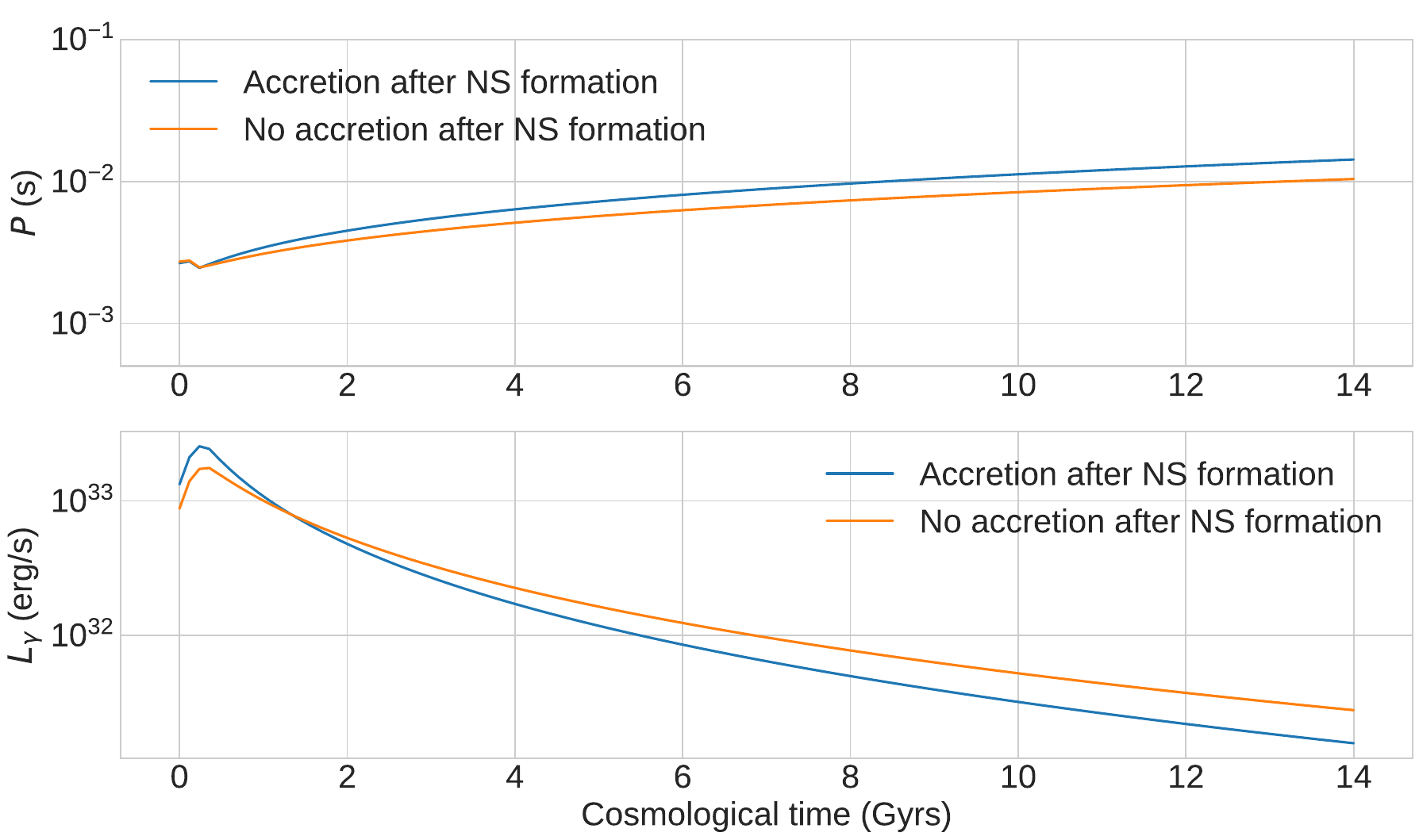}
        % \decoRule
        \caption{{\bf Period and prompt $\gamma$-ray luminosity evolution of a typical AIC NS}. Two cases are compared: where the NS is non-accreting or isolated, and where the NS is in a binary and accretes from its companion.}
        \label{fig:AccVSnoAcc}
\end{figure}

In this section we analyse the effects of the different spin evolution processes described above. We show in  \autoref{fig:AccVSnoAcc} the impact that turning off accretion after NS formation has on the spin period evolution of a binary.
We find that, while accretion effects may cause some initial enhancement in the $\gamma$-ray emission post AIC, in the long term they actually keep the MSP somewhat dimmer than if no post-AIC accretion occurs.

\begin{figure}[ht]
        \centering
        \adjustimage{width=0.8\linewidth}{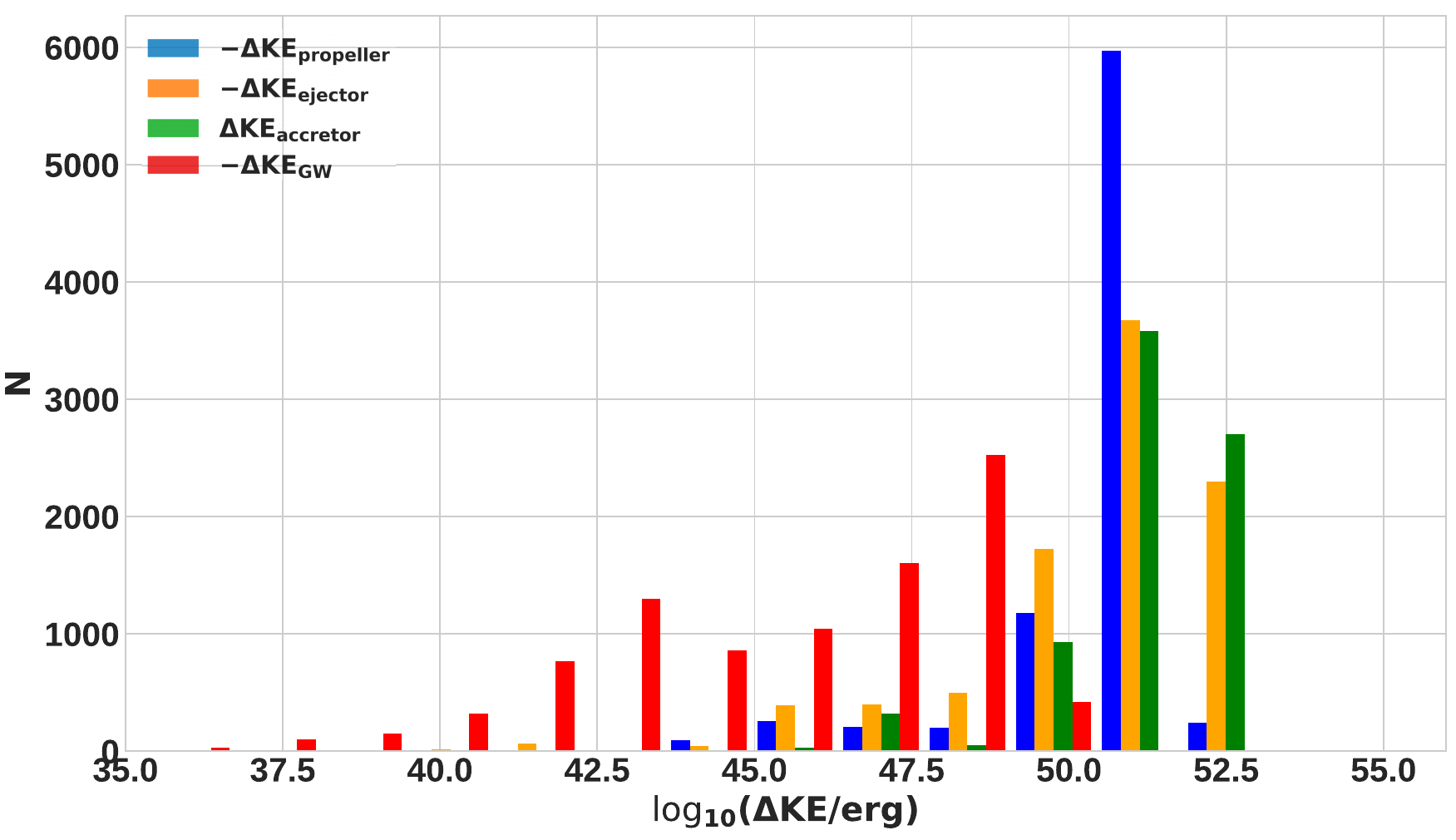}
        % \decoRule
        \caption{{\bf Energy loss/gain mechanisms.} Here we show a distributional comparison of rotational kinetic energy change due to propeller braking,
        magnetic dipole braking (in ``ejector'' phases), accretion spin-up, and gravitational wave emission  for the entire simulated AIC pulsar population (born from a total mass of $2\times 10^9 M_\odot$ evolved with MW Bulge star formation history).}
        \label{fig:delta_KE}
\end{figure}

We show in \autoref{fig:delta_KE} the histograms of the total changes in rotational kinetic energy caused by the various processes outlined in \autoref{sec:spin} up to the end of our 14 Gyr simulation for all model AIC MSPs.
Note that some systems are considerably spun up by accretion and that gravitational wave energy losses at the time of AIC are always small in comparison with all other processes (which act over multi-Gyr timescales).

\section{Further characteristics of the model bulge MSP population}
\label{sec:bulgeMSPs}

\begin{figure}[ht]
        \centering
        %\adjustimage{width=0.8\linewidth}{newFigures/ppdot.pdf}
        \adjustimage{width=0.8\linewidth}{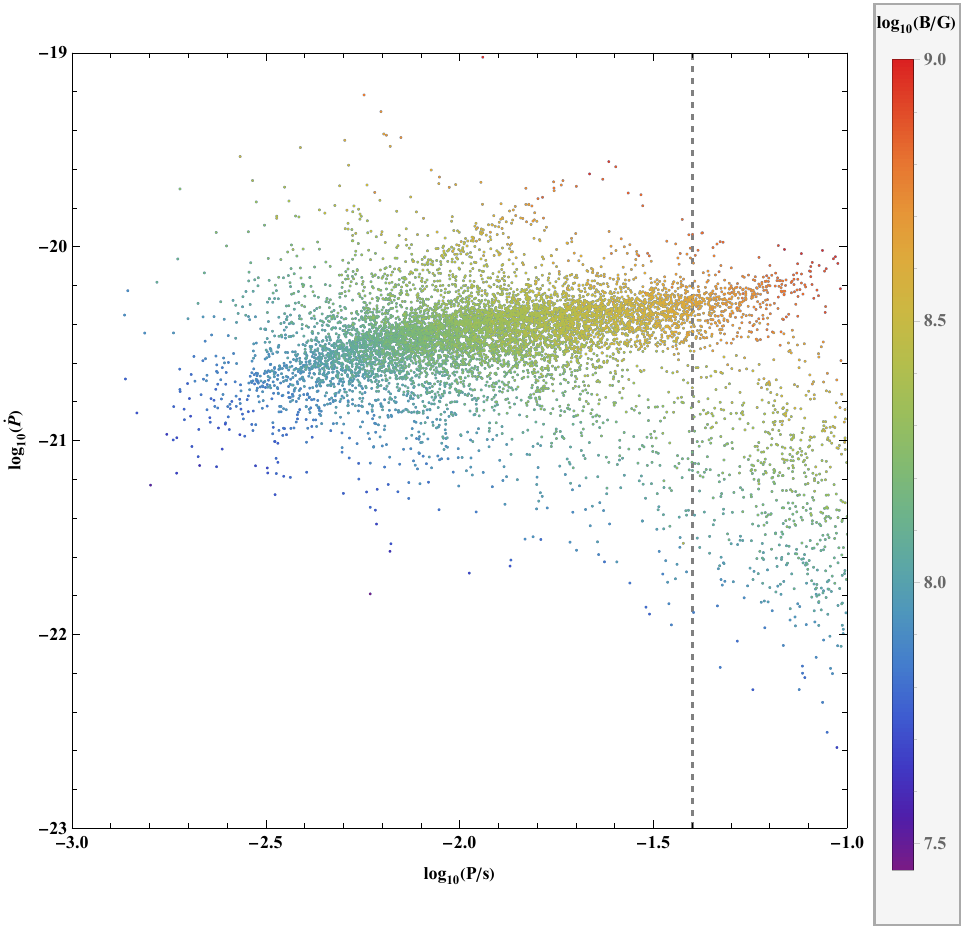}
        % \decoRule
        \caption{{\bf $P-\dot{P}$ diagram of the model Galatic bulge AIC pulsar population} at a cosmological time of 13.8 Gyr. The colour bar displays the magnetic field amplitudes of the pulsars. 
        To the right of the vertical dashed line
       model pulsars have periods longer than 40 ms and are not, therefore, counted as MSPs.}
        \label{fig:ppdot}
\end{figure}

In \autoref{fig:ppdot} we show the $P-\dot{P}$ diagram of our pulsar population at t = 13.8 Gyrs simulation time; colours depict the NSs' magnetic field amplitudes.
The period and prompt $\gamma$-ray luminosities of the model bulge AIC MSP population are shown in \autoref{fig:P_Bulge} and \autoref{fig:L_dist}.
\begin{figure}[ht]
        \centering
        \adjustimage{width=0.8\linewidth}{newFigures/P_dist_Bulge.pdf}
        % \decoRule
        \caption{{\bf Period distribution of the model population of AIC MSPs}. This population corresponds to that expected to be born from a total mass of $2\times 10^9 M_\odot$ evolved with Galactic bulge star formation history. The period distribution is displayed at different times as indicated by the colour scale. }
        \label{fig:P_Bulge}
\end{figure}
\begin{figure}[ht]
            \centering
            \adjustimage{width=0.8\columnwidth}{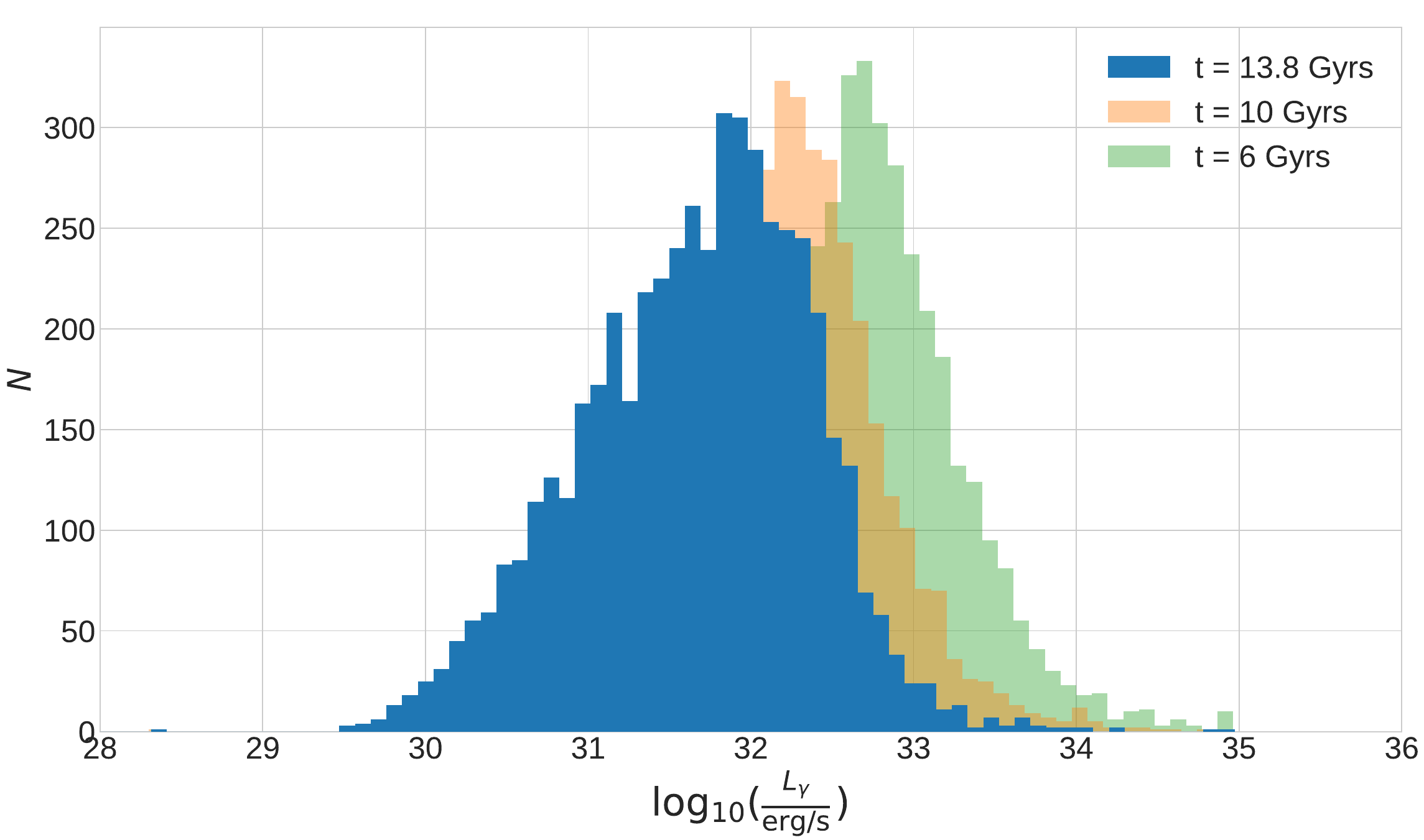}
            % \decoRule
            \caption{{\bf Prompt $\gamma$-ray luminosity distribution of the model population of AIC MSPs}. This population corresponds to that expected to be born from a total mass of $2\times 10^9 M_\odot$ evolved with Galactic bulge star formation history.  The luminosity distribution is displayed at different times as indicated by the colour scale. }
            \label{fig:L_dist}
\end{figure}
\autoref{fig:P_Bulge} and \autoref{fig:L_dist} reveal the slow spin-down and dimming experienced by the bulge population of MSPs over many Gyrs.

\section{Initial calculation of model MSP population prompt flux}
\label{sec:MCassignments}

Our aim is to determine the $\gamma$-ray spectrum received at the Earth due to our model population of bulge MSPs to assess whether this is compatible with the observed GCE.
In our BPS modeling we evolve the spin period of a population of AIC NSs over cosmological timescales.
The (time dependent) masses of these NSs are also determined within the model.
The magnetic field strength of each NS is chosen by sampling from the magnetic field distribution recently determined by ref.~\cite{Ploeg2020}. 
This, in turn, allows us to determine the period derivative attributable to magnetic dipole braking, $\dot{P}_{\rm MB}$.
With $P$ and $\dot{P}_{\rm MB}$ determined for each NS, we can calculate the instantaneous
spin-down power liberated by magnetic braking as:
\begin{equation}
\dot{E}_{\rm MB} =  \frac{4 \pi^2 \ I}{P^3} \dot{P}_{\rm MB} \, ;
\label{eq:Edot}
\end{equation}
note here that $\dot{E}_{\rm MB}(t)$ corresponds to the blue band in Fig.~1.

Next, to determine the  prompt  luminosity of each model MSP and the resulting total prompt signal,
we apply Eq.~1 from the main text  together with measurements of the numerical coefficients obtained by \cite{Ploeg2020} in a combined statistical analysis of data for nearby, resolved $\gamma$-ray MSPs and the GCE signal.
Specifically, we draw  $\log_{10}(f_{\rm \gamma, prompt})$ from a normal distribution with mean 41.4 and standard deviation  $0.6$ and we set  $\alpha_\gamma = -1.9$ and $\beta_\gamma = 0.69$.
We then assume the following spectral form for the prompt $\gamma$-ray spectrum of the $i$th MSP \cite{Ploeg2020}:
\begin{equation}
    \frac{dN_{\rm \gamma,prompt,i}}{dE_\gamma} = K_i E_\gamma^{-\Gamma_i} e^{(-E_\gamma/E_{\rm cut,i})^{2/3}} \, ,
\label{eq:Lgamma_spectrum}
\end{equation}
where, $K_i$ is a proportionality constant determined from: 
\begin{equation}
    \int_{0.1 \rm{GeV}}^{100 \rm{GeV}} E_\gamma \frac{dN_{\rm \gamma,prompt,i}}{dE_\gamma} dE_\gamma = F_{\rm \gamma,prompt,i} \equiv
    \frac{L_{\rm \gamma,prompt,i}}{4 \pi d_i^2} \,.
\label{eq:Knorm}
\end{equation}
In \autoref{eq:Lgamma_spectrum} $E_{\rm cut,i}$ and $\Gamma_i$ are assigned to each MSP 
by drawing from normal distributions with mean and standard deviation determined by \cite{Ploeg2020}.
Specifically,  $\log_{10}(E_{\rm cut} / \text{MeV})$ and $\Gamma$ are drawn from a bivariate normal distribution with a spin-down power-dependent mean and a correlation coefficient. The distribution of $\log_{10}(E_{\rm cut})$ has a mean at $a_{E_{\rm cut}} \log_{10}(\dot{E}/(10^{34.5} \textrm{ erg s}^{-1})) + b_{E_{\rm cut}}$ and a standard deviation $\sigma_{E_{\rm cut}}$. For $\Gamma$, the mean is at $a_{\Gamma} \log_{10}(\dot{E}/(10^{34.5} \textrm{ erg s}^{-1})) + b_{\Gamma}$ and the standard deviation is $\sigma_{\Gamma}$. Here we use $a_{E_{\rm cut}} = 0.23$, $b_{E_{\rm cut}} =2.99$, $\sigma_{E_{\rm cut}} = 0.22$, $a_{\Gamma} = 0.48$, $b_{\Gamma} =0.99$, $\sigma_{\Gamma} = 0.32$ with a correlation coefficient between $\Gamma$ and $E_{\rm cut}$ given by $r_{\Gamma, E_{\rm cut}}=0.73$. These parameters were the highest likelihood parameters in ref.~\cite{Ploeg2020} for a randomly selected Markov chain (see their model A6). The functional form for the prompt luminosity is that of ref.~\cite{Gonthier2018}. 
The error in the prompt flux estimation arising from the uncertainties in these parameters is discussed in S.I. section~\ref{sec:errors}.

With the prompt luminosity and the spectral distribution of the prompt emission of each MSP determined, we then assign, as described by \cite{Ploeg2020}, a distance to each MSP $d_i$ by Monte Carlo sampling from a detailed spatial model of the ``Boxy Bulge'' \cite{Freudenreich1998,Macias2019}. 
The weighted mean distance that we obtain with this procedure is 
$\langle d_{\rm MSP} \rangle \equiv \left(\Sigma_i L_i \Sigma_j d_j^2/L_j \right)^{1/2}= 7.89$\,kpc.
Furthermore, once we have assigned a 3D position to each MSP, we can also derive their Galactic longitudes and latitudes which are required because the flux sensitivity of {\it Fermi}-LAT is dependent on pointing direction (see \autoref{sec:probdetect}).

To recapitulate: We compute the prompt spectral luminosity of each  MSP via Eq.~1 from the main text and solve for $F_{\rm \gamma,prompt,i}$, thence $K_i$, to obtain the prompt differential flux (\autoref{eq:Lgamma_spectrum}) for each MSP.
The predicted, current-day synthetic flux density ($E_\gamma^2$-weighted) is acquired by adding the contribution from all MSPs and is shown as the solid blue line in \autoref{fig:L_specOLD} together with the observed spectrum of the GCE \cite{Macias2019}.

\section{Determining Inverse Compton and Total Spectra}
\label{ref:ICspectrum}

Unfortunately, unlike the case presented by the prompt emission, the amplitude and spectrum of the delayed IC emission from each MSP is subject to many uncertainties.
These include the efficiency with which magnetosphere-escaping $e^\pm$ are reaccelerated at MSP wind shocks and/or intra-binary shocks formed between the pulsar wind and a close companion's stellar wind, and the relative energy densities of ISM magnetic field and radiation field 
(according to which $e^\pm$ energy losses are apportioned into either synchrotron or IC radiation). 
On the other hand, this signal is significantly constrained (as explained below) and we may therefore estimate it using a fitting procedure.
Physically, our model of the IC component assumes emission from a power-law cosmic ray $e^\pm$ population, with an exponential cut-off, IC-scattering a photon population representative \cite{Ackermann2014} of the inner Galaxy.
The normalisation, spectral index, and cut-off energy of the overall $e^\pm$ population are variables in the fit.

\subsection{Initial spectrum determination}

\begin{figure}[ht]
            \adjustimage{width=1\columnwidth}{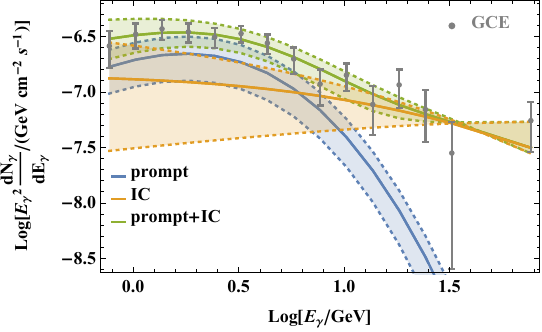}
            % \decoRule
            \caption{{\bf Figure 2 | (Preliminary) GCE spectrum at the Earth.}
            Data points display the measured GCE spectrum
            from ref.~\cite{Macias2019}.
            The curves are as follows:-
            {\bf blue:}
            Bulge prompt $\gamma$-ray (blue) flux 
            at Earth of the model MSP population at 13.8\,Gyr
             predicted using Eq.~1 from the main text and adopting central values of bulge stellar mass, etc. with  1-$\sigma$ error band encompassing 
            the overall fractional error of $1\pm0.52$ on the prompt flux explained in  S.I. section \ref{sec:errors};
            {\bf yellow:}  IC $\gamma$-ray emission;
            and {\bf green:} total $\gamma$-ray emission (green).
            The latter two preliminary curves have been obtained by fitting to the residuals obtained by subtracting the predicted prompt emission from the measured \cite{Macias2019} GCE data points.
            The error bands on the IC and prompt + IC curves encompass the fits obtained for residuals calculated when assuming values of the predicted prompt emission at the top and bottom of the blue band.
            }\label{fig:L_specOLD}
\end{figure}

Given the (preliminary) prompt spectrum is now  prescribed up to some error (see S.I.~\ref{sec:errors}),
our first approach to determining the IC spectrum is simply to find the residuals obtained from subtracting the prompt from the measured spectral data points, then  fit these residuals via an IC emission model; cf.~\autoref{fig:L_specOLD}.

The normalisation, spectral index, and cut-off energy of the overall $e^\pm$ population are variables in the fit; for the latter two we find best-fit values via this exercise of 
$\gamma_e = -2.9_{-0.4}^{+0.3}$ and 
$\log_{10}({E_{\rm e,cut}/{\rm eV}}) = 12.1_{-0.1}^{+1.3}$.
Here the error ranges are given by the best-fits to the residuals defined by adopting extremal values of the prompt flux, i.e., the top and bottom of the blue error band on the prompt spectrum displayed in \autoref{fig:L_specOLD}. 
These fit values are physically plausible and thus entirely consistent with a putative MSP origin of the IC-emitting $e^\pm$ (e.g.,\cite{Harding2021}).
Fractionally, the overall luminosity ($>0.1$ GeV) is $0.46^{+ 0.26}_{ - 0.19}$ due to prompt emission and the total $\gamma$-ray luminosity is $2.3^{+3.1}_{- 1.1} \times 10^{37}$ erg/s
(cf.~the spin-down power liberated by magnetic dipole braking at 13.8 Gyr according to our BPS of $5.8 \pm 0.6 \times 10^{37}$ erg/s) and, formally, we have found a statistically acceptable fit with $\chi^2/dof = 5.1/(15-3) = 0.43$.

\subsection{Constraints on IC and total $\gamma$-ray emission}
\label{sec:ICconstraint}

Despite achieving a good fit, we note that we should consider and exploit, where warranted, a number of physical constraints on the fitted IC and total $\gamma$-ray emission that we have, thus far, ignored.
The implications of these, both singularly and in combination, we now assess in turn. 
They will help inform the final spectral decomposition between prompt and IC emission that we present in the main paper in Fig.~2.

Firstly, energy conservation requires that the luminosity of the combined prompt and IC signals be less than the spin-down power liberated by magnetic braking.
Given the numbers quoted above and with reference to Fig.~1, it is clear that there is sufficient magnetic braking power to energise the total $\gamma$-ray spectrum, although we come interestingly close in the best-fit parameter region to saturating this (cf.~\cite{Sudoh2020}).

Secondly, IC $\gamma$-rays at 1 GeV are up-scattered from the dominant $\sim$ 1.5 eV ISRF of the bulge by 
parent $\sim$ 13 GeV CR $e^\pm$'s.
It is unavoidable that such pairs will synchrotron radiate on the bulge magnetic field.
This has two important consequences: i) the synchrotron luminosity of the model bulge MSP population must also be accounted for in the overall energetics and ii) we must assess whether the predicted synchrotron emission obeys empirical constraints derived from microwave frequency observations of the bulge (see below).

\begin{figure}[ht]
        \adjustimage{width=0.8\linewidth}{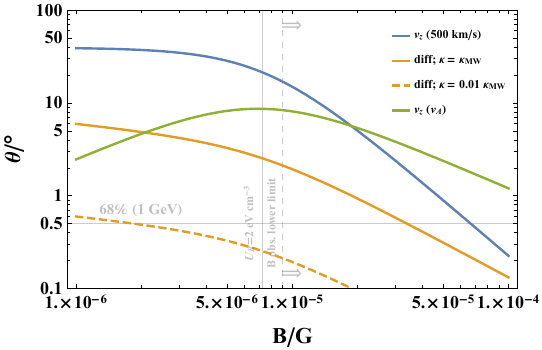}
        % \decoRule
        \caption{{\bf Expected angular transport scales vs. magnetic field amplitude for CR $e^\pm$ IC-radiating 1 GeV $\gamma$-rays} off the ISRF of the bulge where $U_{\rm ISRF, bulge} = 2$ eV cm$^{-3}$\cite{Ackermann2014}. 
        The blue curves are for advective transport at 500 km/s nuclear wind, the yellow for diffusive transport with: i) a CR diffusion 
        similar to that in plane of the Galaxy \cite{Gabici2007}) and ii) suppressed by $0.01$ with respect to this value; and green lines are for transport at the Alfv\'en speed (i.e., for streaming along the
        field lines of the bulge magnetic field).
        The horizontal line at $0.5^\circ$ shows the 68\% angular containment radius of the {\it Fermi}-LAT at 1 GeV \cite{Atwood2009}.
The vertical dashed line indicates the lower limit on the magnetic field amplitude in this region of the Galaxy established in \cite{Carretti2013}.
        }
        \label{fig:plot1GeVtransport}
\end{figure}

Thirdly, whilst one might seek to evade these synchrotron constraints by dialing down the bulge magnetic field amplitude (which we now treat as a free parameter to be determined), this strategy does not work because 
i) there is an empirical lower limit on the magnetic field amplitude of $\sim 9 \ \mu$G in this region \cite{Carretti2013}
and ii) the IC signal would then become too smeared out.
In more detail about the latter:
because the putative IC-radiating secondary CR $e^\pm$ pairs are launched into the ISM, this introduces the possibility that, unlike the prompt signal which emerges from the magnetospheres of individual MSPs, the delayed signal is smeared out over some angular scale corresponding to the distance over which the pairs are transported during their energy-loss timescale.
We show the equivalent one-dimensional smearing angular scale for CR $e^\pm$ IC-radiating at 1 GeV (and losing energy via their synchrotron and IC emission) as a function of bulge magnetic field in
\autoref{fig:plot1GeVtransport}.
These curves have been evaluated for a bulge ISRF energy density of $\sim$ 2 eV cm$^{-3}$\cite{Ackermann2014}.
The different colours illustrate the results obtained for different assumptions about the transport of radiating pairs in the bulge.

\section{Broadband modeling}
\label{sec:BB}

To gain further insights and to account mathematically for the constraints described above, we now perform a classical $\chi^2$ analysis wherein we allow the normalization of the prompt spectrum to float (keeping the spectral shape fixed) in addition to the normalization, spectral index, and cut-off energy of the IC-emitting CR $e^\pm$ spectrum.
Where required, we also introduce a floating bulge magnetic field amplitude.

\begin{figure}[ht]
        \adjustimage{width=0.7\linewidth}{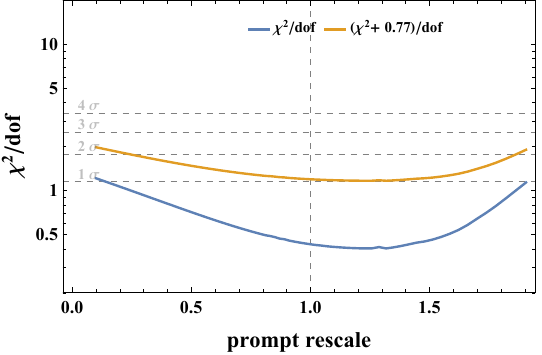}
        % \decoRule
        \caption{{\bf $\chi^2$ per degree of freedom} ($dof = 12$) for a fit of  combined prompt + IC emission to the GCE data \cite{Macias2019} as a function of the rescaling of the amplitude of the prompt flux away from the central value determined in \autoref{sec:MCassignments}. For the yellow curve we renormalize the min of $\chi^2/dof$
        to 1 in order to establish the $\pm 1 \sigma$ confidence interval on the rescaling parameter.}
        \label{fig:plotChi2Anuj}
\end{figure}

In our baseline model, we consider the GCE $\gamma$-ray data points but also introduce a single additional datum that penalizes the $\chi^2$
for renormalization of the prompt signal amplitude away from the central value we have already calculated.
From this we find a minimum $\chi^2/dof$ of $\sim$0.4 and a $1\sigma$ range for the prompt rescaling of $\sim 0.3-1.9$; cf.~\autoref{fig:plotChi2Anuj}.
Note that this preliminary fit does not require either that the 
implied synchrotron flux (for the minimum empirically acceptable magnetic field of $9 \ \mu$G) obey the empirical  upper limits that emerge from microwave observations \cite{Ade2013} nor that the implied total luminosity be less than the spin down power determined from our BPS.
We find, indeed, that over much of the range of the rescaling parameter within  the 1$\sigma$ confidence region determined above one or both of these constraints is violated.

\subsection{MSP contribution to the Microwave Haze}

\begin{figure}[ht]
        \adjustimage{width=\linewidth}{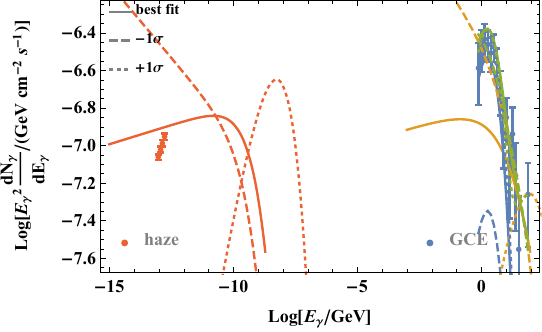}
        % \decoRule
        \caption{{\bf Spectral energy distribution from radio to $\gamma$-ray energies for the model bulge MSP population} adopting best-fit parameters determine in \autoref{sec:BB} and a magnetic field amplitude 9 $\mu$G (the empirical lower limit on the bulge field \cite{Carretti2013}). Curves are as follows:- blue: prompt emission; yellow: IC; blue: total $\gamma$-ray emission; and red: synchrotron.
        The solid curves are evaluated for best-fit parameters, dashed curves evaluated for parameters at the lower edge of the $1 \sigma$ confidence region (prompt rescaling $\sim 0.2$) and dotted curves are evaluated for parameters at the top edge (prompt rescaling $\sim 1.9$).
        }
        \label{fig:plotBB1}
\end{figure}

\begin{figure}[ht]
        \adjustimage{width=\linewidth}{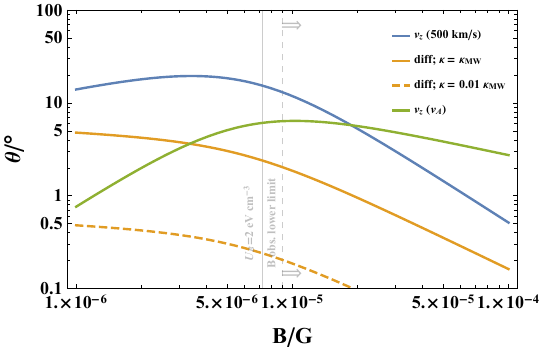}
        % \decoRule
        \caption{{\bf Expected angular transport scales vs. magnetic field amplitude for CR $e^\pm$ synchrotron-radiating at 44\,GHz} off a bulge magnetic field with the nominated amplitude and accounting for IC losses on a $U_{\rm ISRF, bulge} = 2$ eV cm$^{-3}$ ISRF. The blue curves are for advective transport at 500 km/s, the yellow for diffusive transport (with CR diffusion coefficient suppressed by $\chi = 0.01$ with respect to Galactic plane values as motivated by observation of TeV halos around local pulsars), and green lines are for transport at the ion Alfv\'en speed (i.e., for streaming along 
       magnetic  field lines) for $n_H = 0.01$ cm$^{-3}$. The vertical dashed line indicates the lower limit on the magnetic field amplitude in this region of the Galaxy established in \cite{Carretti2013}.}
        \label{fig:plotHazetransport}
\end{figure}

\begin{figure}[ht]
        \adjustimage{width=0.8\linewidth}{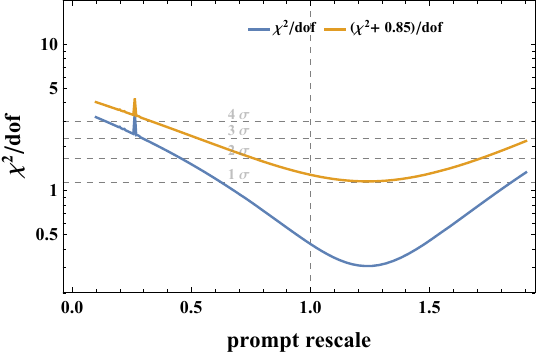}
        % \decoRule
        \caption{{\bf $\chi^2$ per degree of freedom} ($dof = 16$) for a fit of  prompt + IC emission to the GCE data \cite{Macias2019} combined with a synchrotron to the Microwave Haze data \cite{Ade2013}
         as a function of the rescaling of the amplitude of the prompt flux away from the central value determined in \autoref{sec:MCassignments}. For the yellow curve we renormalize the min of $\chi^2/dof$
        to 1 in order to establish the $\pm 1 \sigma$ confidence interval on the rescaling parameter.}
        \label{fig:plotChi2AnujII}
\end{figure}

\begin{figure}[ht]
        \adjustimage{width=0.8\linewidth}{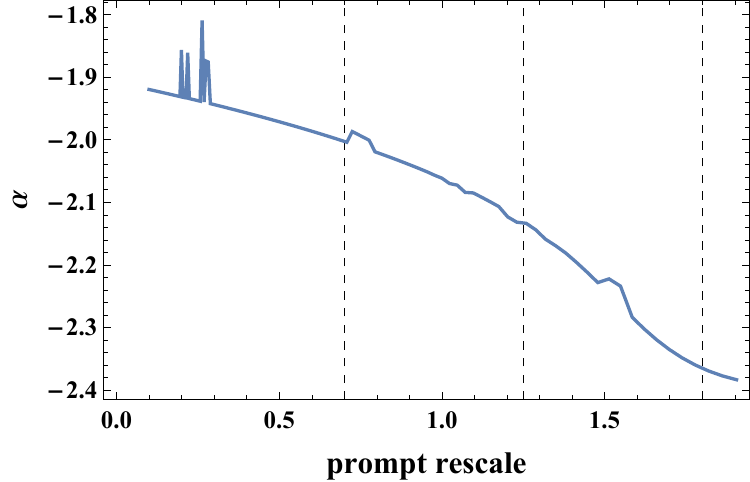}
        % \decoRule
        \caption{{\bf Spectral index of the model CR $e^\pm$ population as a function prompt flux rescaling}. The vertical dashed lines indicate the best fit and $\pm 1 \sigma$ region.
        }
        \label{fig:plotalphaGCE}
\end{figure}

\begin{figure}[h]
        \adjustimage{width=0.8\linewidth}{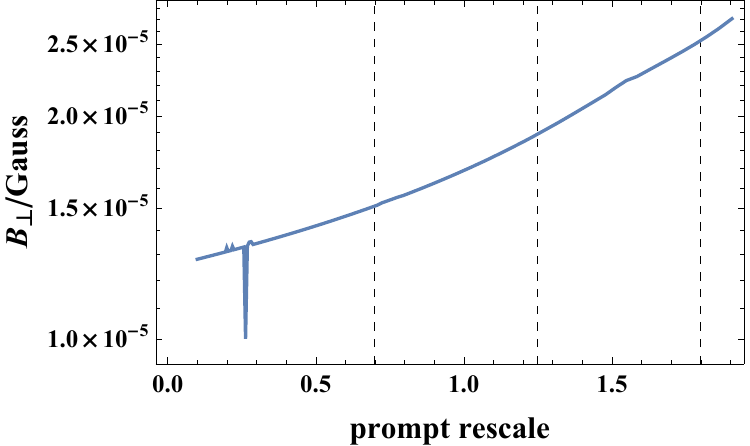}
        % \decoRule
        \caption{{\bf Amplitude of the perpendicular component of the magnetic field as a function prompt flux rescaling}. The vertical dashed lines indicate the best fit and $\pm 1 \sigma$ region.
        }
        \label{fig:plotBperpGCE}
\end{figure}

\begin{figure}[h]
        \adjustimage{width=0.8\linewidth}{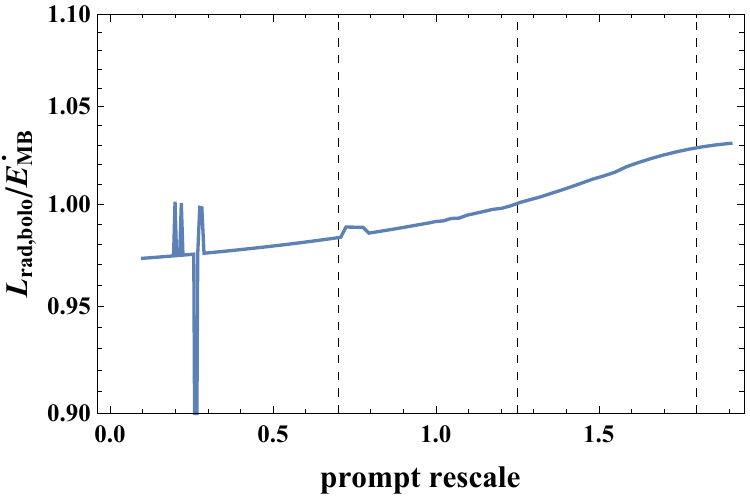}
        % \decoRule
        \caption{{\bf Ratio between the bolometric radiated luminosity (in synchrotron + IC + prompt emission) and the power liberated by magnetic braking at 13.8\,Gyr} from the bulge  MSP population as a function of the prompt flux rescaling. The vertical dashed lines indicate the best fit and $\pm 1 \sigma$ region.
        }
        \label{fig:plotPowerRatio}
\end{figure}

Consider the broadband spectral energy distribution predicted for the best-fit parameters obtained in the preliminary fitting described above: see \autoref{fig:plotBB1}.
What is tantalising about this figure is that while the predicted synchrotron radiation is clearly somewhat in excess of the microwave data measured with the Planck satellite \cite{Ade2013}, it is nevertheless rather similar in amplitude.
These data points describe the spectrum of the Microwave Haze \cite{Finkbeiner2004}.
The haze apparently extends up to somewhat larger angular scales ($\sim 35^\circ$)
than the bulge and a morphological correspondence between it and the even more extended ($\sim 55^\circ$) Fermi Bubbles has been claimed (e.g., \cite{Ade2013}).
On the other hand, in its lower reaches the Haze is certainly coincident with the bulge and, as discussed above, transport effects should carry microwave-emitting 
CR $e^\pm$ over an angular scale of $\sim 5-10^\circ$ after they are launched by bulge MSPs that are themselves distributed out to $\sim 20^\circ$ 
(cf.~\autoref{fig:plotHazetransport}).
It is also, of course, possible that more than one processes or population contributes CR $e^\pm$ that generate the Haze.
In any case, motivated by these considerations, we perform a new $\chi^2$ analysis wherein we introduce the magnetic field amplitude as a further tunable parameter, and we fit the combination of the Haze microwave data, the GCE $\gamma$-ray data, a datum for the prompt emission rescaling factor, and a datum  which penalizes the $\chi^2$ for the difference between the total luminosity and the total spin-down power from magnetic braking at 13.8\,Gyr according to our BPS ($1.0 \pm 0.1 \times 10^{38}$ erg/s).
In other words, we assume here that radiative processes saturate the spin-down power (cf.~\cite{Sudoh2020}).
The results from this are shown in \autoref{fig:plotChi2AnujII} and subsequent figures.
Not unexpectedly, the region of good fit is now somewhat more constrained to the region  $\sim 0.8 -  1.7$.
The broadband spectral energy distribution obtained for the new best-fit parameters is shown in the main text as
Fig.~3.
%\autoref{fig:BB_spec}.
%
Three other quantities of interest -- the spectral index of the model CR $e^\pm$ population, the amplitude of the perpendicular component of the bulge magnetic field, and the fraction of spin-down power liberated by magnetic braking that is radiated in non-thermal emission -- are shown as functions of the rescaling in  \autoref{fig:plotalphaGCE}, 
\autoref{fig:plotBperpGCE}, and \autoref{fig:plotPowerRatio}.
\begin{figure}[h]
        \adjustimage{width=0.8\linewidth}{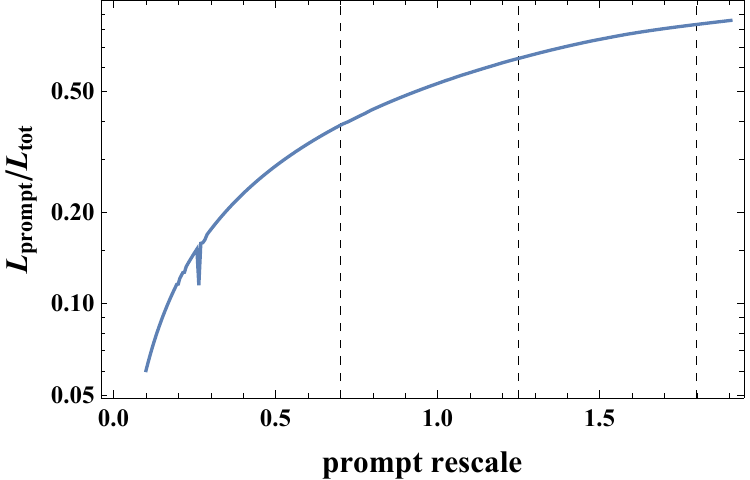}
        % \decoRule
        \caption{{\bf Ratio between the bolometric power radiated into prompt emission to all radiation} (synchrotron + IC + prompt emission)  from the bulge  MSP population at cosmological time 13.8\,Gyr as a function of the prompt flux rescaling. The vertical dashed lines indicate the best fit and $\pm 1 \sigma$ region.
        }
        \label{fig:plotPromptToTotRatio}
\end{figure}
These figures demonstrate the overall self-consistency of a scenario 
that invokes a combination of both IC and prompt emission from MSPs as an explanation of the GCE while the CR $e^\pm$ launched into the ISM by the MSPs and requisite to explain the observed IC emission also explain the observed microwave Haze.
One question hanging over this analysis concerns the very hard spectral index, $\sim -2.1-2.2$ of the steady state CR $e^\pm$ (required to simultaneously explain haze and GCE).
For reasonable transport and ISM parameters, this population has to largely cool
{\it in situ} implying, in the case of a power law {\it at injection},
an injection spectral index of $\sim 1.1-1.2$.
Alternatively, a monochromatic or highly peaked injection spectrum would also work.
Neither of these is excluded by current research into MSPs (see the discussion in \cite{Sudoh2020}; we leave further consideration of this issue to subsequent work).
In any case, the best-fit parameters found in this analysis are those we adopt in generating the GCE spectral decomposition illustrated in Fig.~2
of the main text.

\section{Error budget(s)}
\label{sec:errors}

The two dominant contributions to the total error in rescaling our results
(for the number of AIC systems) from our BPS modeling of $\sim 6.3 \times 10^7$ binary systems (equivalent host stellar mass of $2.0 \times 10^9 M_\odot$) to the Galactic bulge are due to i) the uncertainty in the current stellar mass of the bulge and ii) the uncertainty in the bulge binarity fraction.
For the former, ref.~\cite{Portail2015} states that the bulge stellar mass is $(1.4-1.7) \times 10^{10} M_\odot$ which we treat as $(1.55 \pm 0.15) \times 10^{10} M_\odot$ (1$\sigma$). 
For the latter, if we allow that the bulge binarity fraction may range over the (likely extremal) range $50-100\%$, then the possible  fractional change in ZAMS mass of the parent stellar population relative to our fiducial assumption of 70\% binarity fraction is $\sim 1.0 \pm 0.27$.
Added in quadrature, the overall fractional error accruing in scaling from the number of binaries we have modeled to that appropriate for bulge stellar population is $1 \pm 0.30$.
Other sources of error arising from systematic uncertainties in accretion physics on the number of AIC events, or the correct parameter choices to describe this or, e.g., assumptions about initial eccentricity distributions
are discussed in ref.~\cite{Ruiter2019}, section 4.1; these are small in comparison with the uncertainty accruing from the binarity fraction. 

In addition to these two uncertainties, a further significant uncertainty is encountered in assessing the expected prompt $\gamma$-ray luminosity of the model MSP population.
This stems from uncertainties in the physical values of the numerical coefficients entering into Eq.~1 from the main text --$f_{\rm \gamma,prompt},\alpha_\gamma,\beta_\gamma$ -- which have been estimated on the basis of a combined statistical analysis in ref.~\cite{Ploeg2020} of local, resolved MSPs and the GCE signal.
In order to assess the scale of this uncertainty we generate the distribution of prompt spectra by simulating 1000 bulge populations of MSPs. 
For each population we sample at random a parameter set from the Markov chains produced for model A6 in \cite{Ploeg2020} and assign each pulsar a luminosity, spectral index, $E_{\rm cut}$ depending on the MSP's period and the magnetic dipole braking component of the period derivative. %
We also simulate a position by sampling from the Boxy Bulge density model. 
The simulated spectrum for the prompt emission of each population is then the sum of the MSP spectra within the region of interest.
We determine the bin-by-bin means and standard deviations of this collection of MSP population spectra. 
From this exercise, the flux-weighted fractional error averaged over energy bins is $\pm$ 0.43.
We add this in quadrature to the errors due to the binarity fraction uncertainty and the present-day bulge stellar mass uncertainty to obtain a total (fractional) prompt $\gamma$-ray luminosity error of  $\pm$0.52.

\section{MSP detection probability}
\label{sec:probdetect}

The \textit{Fermi}-LAT has a point source flux sensitivity of around
$F_{>0.1\rm GeV} \sim  10^{-12} \ \rm erg\ cm^{-2}\ s^{-1} $,
but this varies across the sky and it is, in general, more difficult to positively identify a $\gamma$-ray pulsar than a generic $\gamma$-ray point source \cite{Abdo2013}.
We use the Monte Carlo procedure developed by \cite{Ploeg2020} to estimate the flux for each model MSP such that, given its assigned sky position 
and spectrum, it could be  resolved as a $\gamma-$ray point source ($F_{\rm sens,point}$) and possibly as an individual MSP ($F_{\rm sens,MSP}$)

Following \cite{Ploeg2020} and references therein, we assume that the MSP detection threshold sensitivity flux $F_{\rm sens}$ is drawn from a log-normal distribution (see eq.~2.34 of \cite{Ploeg2020}).
Concerning the threshold sensitivity flux, $F_{\rm sens, MSP}$, for detection and positive identification of model MSPs as MSPs we assume that  $\log_{10}(F_{\rm sens, MSP})$ is distributed according to a normal distribution with mean $\log_{10} (\mu_{\rm th}) + K_{\rm th}$ and standard deviation $\sigma_{\rm th}$.
Here $\mu_{\rm th}$ is the longitude and latitude dependent {\it Fermi}-LAT point source flux detection threshold ($\mu_{\rm th}(l,b)$ is associated with the 4 FGL-DR2 catalog; see \url{https://fermi.gsfc.nasa.gov/ssc/data/access/lat/10yr_catalog/}), and $K_{\rm th} = 0.458904$ and $\sigma_{\rm th} = 0.251911$ are parameters determined in \cite{Ploeg2020}, invariant across the MSP population; $K_{\rm th}$ accounts for the difficulty inherent in identifying a pulsar as distinct from a $\gamma$-ray point source (so for calculations related to the probability of detecting models MSPs as resolved $\gamma$-ray point sources of undetermined type we set $K_{\rm th} = 0$). 

\section{Model MSP population (sub-threshold) flux distribution}
\label{sec:subthresh}

\subsection{Population of $\gamma-$ray resolvable Bulge MSPs}

\begin{figure}[h]
        \adjustimage{width=0.8\linewidth}{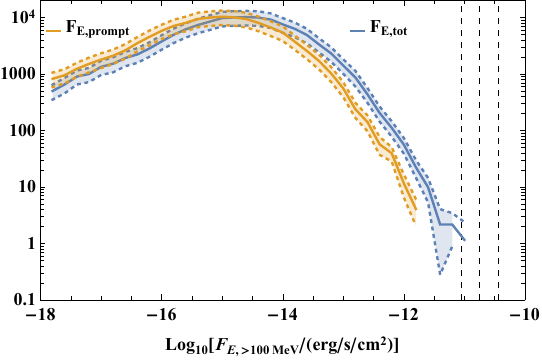}
        % \decoRule
        \caption{{\bf Total and prompt energy flux distributions of the modeled MSP population at $t = 13.8$\,Gyrs cosmological time.} 
        The vertical dashed lines show the mean and $\pm 1\sigma$
        point-source detection thresholds \cite{Ploeg2020}.
        Data are histogrammed into bins of width $\Delta$log$_{10}\left[F/\left({\rm erg \ s^{-1} \ cm}^{-2}\right)\right]$ = 0.2.
        Error bands encompass 1-$\sigma$ errors.
        }
        \label{fig:Fdist1}
\end{figure}

In \autoref{fig:Fdist1}  we show the absolute (energy) flux $F_E$ distribution of our model bulge AIC MSP population at 13.8 Gyr.
The middle panel shows
a zoom-in, close to the detection threshold (0 on the x-axis),
of the histogrammed  distribution of the ratios of each MSP's prompt flux to the  flux required for likely detection, 
either as a $\gamma-$ray point source ($F_{\rm sens,point}$) or (more stringently) as an individual MSP ($F_{\rm sens,MSP}$),
at the MSP's modelled position (S.I.~\ref{sec:probdetect}).
Very few MSPs will be detected individually.
Most of the CR $e^\pm$ launched by each MSP will be transported outside the aperture defined by the \fermi \ point spread function before IC radiating (cf.~\autoref{fig:plot1GeVtransport}), so the IC flux contributed by each MSP does not contribute to its point source flux.
\begin{figure}[h]
        \adjustimage{width=0.8\linewidth}{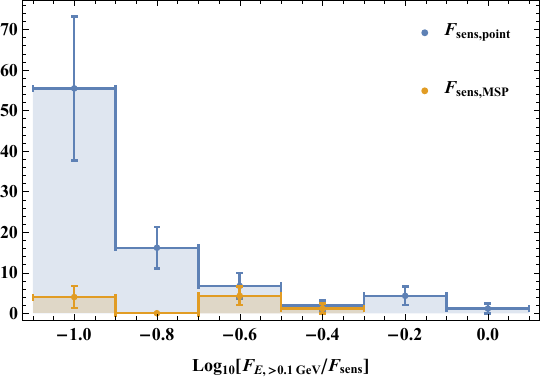}
        % \decoRule
        \caption{{\bf Zoom-in (close to the flux detection thresholds) of the
        MSP population prompt energy fluxes}
         normalised to the {\it Fermi}-LAT detection threshold for i) MSPs and ii) $\gamma$-ray point sources.
         Horizontal bars show the (logarithmic) bins widths;
         vertical errors encompass 1-$\sigma$ errors.
        }
        \label{fig:Fdist2}
\end{figure}
\begin{figure}[h]
        \adjustimage{width=0.8\linewidth}{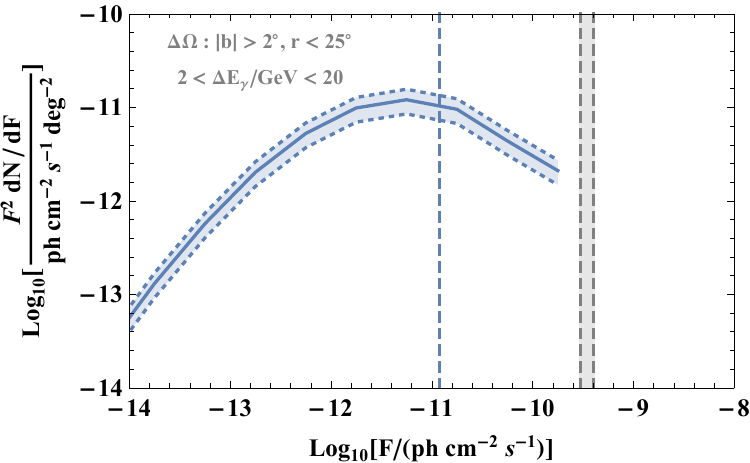}
        % \decoRule
        \caption{{\bf The number flux distribution of the model bulge MSP population} 
         over the solid angle defined by $|b| > 2^\circ$ and $r < 25^\circ$;
the vertical, dashed, blue line is the number flux equivalent to an expected 1 photon per source; the vertical, dashed, gray lines subtend the (variable) approximate point source flux sensitivity threshold \cite{Chang2020}.
Note that the population contains no MSPs with $F_E\gtrsim 10^{-11}$ erg/s/cm$^2$ or $F \gtrsim 10^{-10}$ ph/s/cm$^2$.
The (blue) error bands encompass 1-$\sigma$ errors.
        }
        \label{fig:Fdist3}
\end{figure}
The photon number flux distribution of the population is shown in  \autoref{fig:Fdist3}.

\subsection{Confronting our predictions with point source inference}

\begin{figure}[h]
        \adjustimage{width=0.8\linewidth}{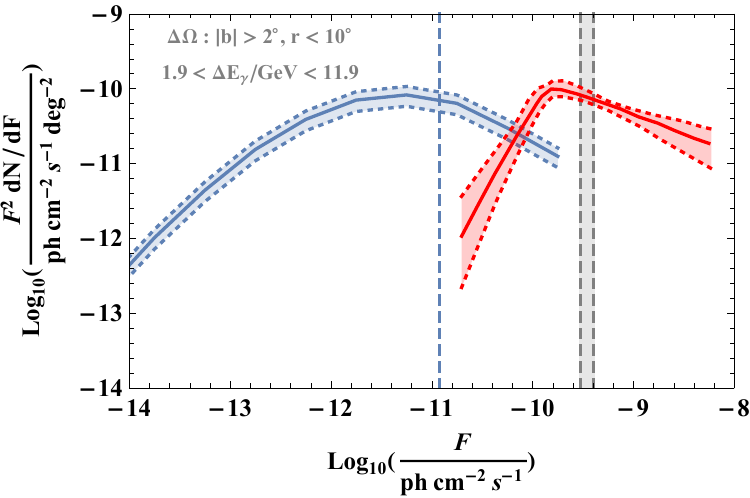}
        % \decoRule
        \caption{{\bf Source count distribution of our model population
for $\gamma$-ray energies 1.9-11.9 GeV} and a normalising solid angle of 234 square degrees (as roughly compatible with \cite{Lee2016}.)
shown as the blue band;
the red band shows the best-fit source-count function for point sources within
the given solid angle 
using non-Poissonian template fitting and with 3FGL sources unmasked as obtained from \cite{Lee2016}.
Both coloured error bands encompass 1-$\sigma$ errors.
The vertical, dashed, blue line shows approximately the flux equivalent to an expected 1 photon per source; the vertical, dashed, gray lines subtend the (variable) approximate point source flux sensitivity threshold \cite{Chang2020}.
        }
        \label{fig:plotSourceCountDistributionLeeCompare}
\end{figure}

\begin{figure}[h]
        \adjustimage{width=0.8\linewidth}{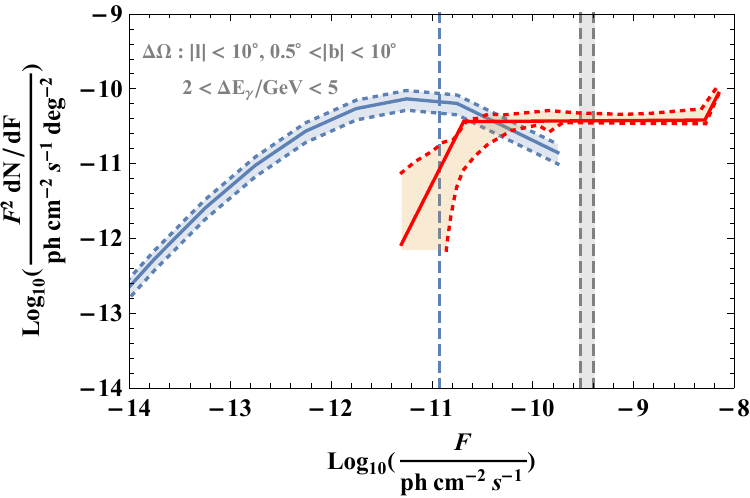}
        % \decoRule
        \caption{{\bf Source count distribution of our model population
for $\gamma$-ray energies 2-5 GeV} and a normalising solid angle of 380 square degrees (as roughly compatible with \cite{Calore2021}.)
The orange band shows the $dN/dF$
distribution obtained by \cite{Calore2021}
within the given solid angle
for the fiducial case
"pPDF-B, sF-noGCE" (in their notation).
Both coloured error bands encompass 1-$\sigma$ errors.
The vertical, dashed, blue line shows approximately the flux equivalent to an expected 1 photon per source; the vertical, dashed, gray lines subtend the (variable) approximate point source flux sensitivity threshold \cite{Chang2020}.
        }
        \label{fig:plotSourceCountDistributionCaloreCompare}
\end{figure}

There are ongoing efforts, labelled collectively as {\it point source inference} \cite{Collin2021}, 
to statistically probe the nature
of sub-threshold and individually-unresolvable point sources putatively contributing to the GCE signal.
To date, 
these efforts have employed a number of different techniques including
non-Poissonian template fitting (``NPTF'';\cite{Lee2016,Chang2020,Buschmann2020,Calore2021}), wavelet analysis \cite{Bartels2016}, and machine learning methods \cite{List2020}.
All of these techniques at least hint at the existence of a significant sub-threshold source contribution to the GCE.
Moreover, some point source inference techniques proffer, at least in principle, the ability not only to evince such sources, but actually to recover their flux distribution, $dN/dF$.
In practice, 
the results emerging from different groups and/or with different techniques have not yet converged
and do not seem to be yet fully robust to systematics
\cite{Leane2019,Chang2020,Leane2020,Leane2020b,Buschmann2020}.
In particular, the ability of such techniques to correctly recover the underlying flux distribution 
of putative GCE sources is affected by our lack of a perfect model of the Galactic foregrounds.
Moreover, the statistical reconstruction of a sub-threshold flux distribution must become increasingly unreliable 
as
the zero photon limit 
is approached 
because such an ``ultra-faint'' source population is degenerate with Poissonian emission \cite{Chang2020}.
The boundary to this region of degeneracy 
is roughly given by the single photon flux
for which only a single photon from a given source is expected over the exposure of the instrument
\cite{Chang2020}.
For {\it Fermi}-LAT GCE observations, this corresponds to
$\sim 10^{-11}$ counts cm$^{-2}$ s$^{-1}$ (e.g., \cite{Chang2020}).
This is
more-or-less where our model MSP distribution peaks (cf.~\autoref{fig:plotSourceCountDistributionLeeCompare})
but given, in particular, that we still predict some brighter sources between the single photon flux
and the point-source sensitivity threshold, the flux distribution we predict will be subject to empirical testing via
point source inference as these techniques mature.
\autoref{fig:plotSourceCountDistributionLeeCompare} and \autoref{fig:plotSourceCountDistributionCaloreCompare}
compare the predictions of our BPS with the results obtained in the pioneering study of \cite{Lee2016}
and the recent study of \cite{Calore2021}.
There is apparently some tension here, but note 
that the results of 
\cite{Lee2016} and \cite{Calore2021}
are themselves discrepant and, what is more,
the level of disagreement between our results
and the point source inference results is
reduced with respect to the more recent study.
Furthermore, while we clearly predict more power in the distribution 
in some flux region
than determined by  \cite{Calore2021}, this 
only occurs for photon fluxes 
in the problematic regime
around and below the single photon limit.
Also note that, because our distribution
cuts off fairly sharply just above a photon flux of $10^{-10}$ cm$^{-2}$ s$^{-1}$, while \cite{Calore2021}
assume a distribution that is continuous to much higher fluxes, 
these results are not strictly comparable.

\section{Predicted Bulge X-ray phenomenology}
\label{sec:Xrays}

Accreting NSs generate X-ray emission. If this is higher than about $10^{36}\ \rm erg/s$, these objects could become detectable as X-ray point sources by INTEGRAL  \cite{Haggard2017}.
The accretion power (the rate at which kinetic energy is dissipated at the stellar surface)
is given by (e.g. \cite{Ruiter2006}) 
\begin{equation}
        L_X = \eta \frac{G M_{acc} \dot{M}_{acc}}{R_{acc}},
 \label{eq:Lxrays}
\end{equation}
where $M_{acc}$ and $R_{acc}$ are the accreting NS mass and radius, respectively, and $\eta$ is the dimensionless, bolometric correction factor which we determine for each system by randomly sampling from a flat distribution ranging from 0.1 to 0.6. 

\begin{figure}[h]
        \adjustimage{width=0.8\linewidth}{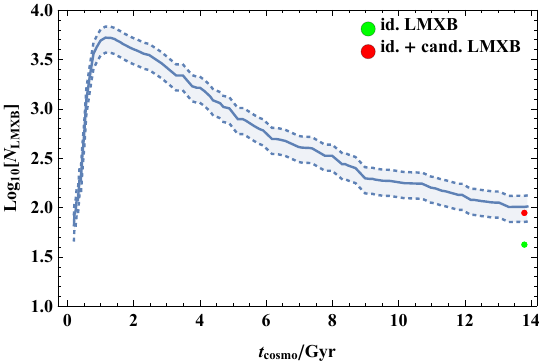}
        % \decoRule
        \caption{{\bf Number of model Bulge INTEGRAL-observable neutron star accretors with $L_X>10^{36} \rm{erg/s}$ 
        vs.~cosmological time}. 
        The blue error band encompasses the 1-$\sigma$ error.
        The data points show the number of identified (green) and identified + candidate (red)
         LMXBs observed \cite{Haggard2017} in the MW Bulge.}
        \label{fig:N_LMXB}
\end{figure}

Using \autoref{eq:Lxrays} we find a population of ${\sim} 6.9 \times 10^3$ neutron star accretors with about 60 having current luminosities higher than $10^{36} \ \rm{erg \ s^{-1}}$  (cf.~\ref{fig:N_LMXB}). This is entirely consistent with INTEGRAL observations \cite{Haggard2017} which have uncovered 42 LMXBs and a further 46 LMXB candidates within $10^\circ$ radius around the Galactic Center.

\begin{figure}[h]
        \centering
        %\adjustimage{width=0.8\columnwidth}{newFigures/L_X_comp.pdf}
        \adjustimage{width=0.8\columnwidth}{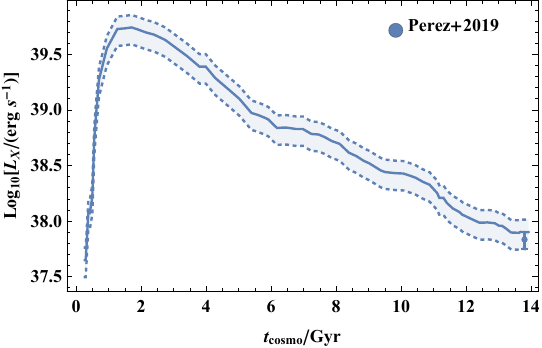}
        % \decoRule
        \caption{{\bf Integrated LMXB X-ray luminosity from the simulated Bulge population over cosmological time.} Shown as a band around the simulated X-ray luminosity time-series is the 1-$\sigma$ error due to Bulge mass uncertainties. The data point shown at t = 13.8 Gyrs is the observed X-ray luminosity from the MW Bulge \cite{Perez2019}.}
        \label{fig:L_x_comp}
\end{figure}

We also need to establish whether the integrated X-ray emission of the unresolved point sources of our model is consistent with observations.
Thus, we use the observed \cite{Perez2019} $2-10$\,keV X-ray luminosity  of the Galactic Bulge ($6.8 \pm 1.2 \times 10^{37} \ \rm erg \ s^{-1} $), shown as the data points in  \autoref{fig:L_x_comp} at 13.8\,Gyrs. 
To calculate the model X-ray luminosity shown in the same figure we use moving average smoothing.
The integrated X-ray luminosity of our model population is, for the central value, close to, but below, that measured, which is totally acceptable considering that other X-ray sources 
may contribute to the total X-ray luminosity of the Bulge.

\section{TeV halos around bulge MSPs?}

An additional argument \cite{Hooper2018,Hooper2018b,Hooper2021} that has been raised
against an MSP origin for the GCE is that, on the basis of HAWC data \cite{Albert2020}
covering a number of local MSPs, one predicts TeV emission in ``halos'' around the putative MSP population in the inner $\sim 0.5^\circ$ that would be in excess of that detected by the HESS instrument \cite{HESS2016}.
Formally, because we model in the main text the Boxy Bulge region and not the Nuclear Bulge (the latter being overlapping with the inner $\sim 0.5^\circ$ field), this constraint does not pertain to our model MSP population.
In any case, with $E_{\rm cut}$ 
typically evaluating to $\sim$ TeV in our analyses -- i.e., with an electron spectrum that cuts off around this energy -- we do not typically predict strong TeV IC emission from MSPs in our model population.
Moreover, as emphasised in the main text, there are systematic differences between the bulge and local, disk MSP populations.
We will return to this issue in a subsequent publication.

\section{Nuclear bulge}
\label{sec:NB}

The spectral $\gamma$-ray data points we plot and model in the main text are those associated \cite{Macias2019} to the Boxy Bulge (BB) \cite{Freudenreich1998} template which encompasses around 90\% of the stellar mass of the overall Galactic bulge.
However, template analyses \cite{Macias2018,Bartels2018,Macias2019} have  demonstrated that a separate stellar structure 
-- the Nuclear Bulge (NB) \cite{Launhardt2002,Bland_Hawthorn_2016} --
makes a distinct, measurable contribution -- at the 15\% level \cite{Macias2019} --  to the overall GCE signal.
The NB covers the central $\sim$\,200 pc around the Galactic centre and has a star formation history distinct from that over the wider-scale bulge \cite{Nogueras-Lara2020}.
This implies that the NB's MSP population's $\gamma$-ray emission must be separately modelled, a task we address here.
In particular, we demonstrate in this section that our BPS model predicts $\gamma$-ray emission from the AIC MSPs consistent with observations.
(Note that the similarity between the fraction of the total bulge stellar mass the NB contributes -- $\sim 10$ \% -- and the fraction of the overall GCE luminosity it seems to generate -- $\sim 15$\% -- is already suggestive that an astrophysical origin for the GCE phenomenon is tenable.)

As previously, our first step is to make a preliminary estimate of the spectrum of prompt emission from the NB.
To achieve this, we create a NB model population of AIC-producing 
binaries by re-drawing from our existing BPS
population adopting the 
NB's
star formation history \cite{Nogueras-Lara2020} 
and scaling for the difference in 
the equivalent
ZAMS mass of our reference data set ($2.0 \times 10^9 M_\odot$) and that of the NB.
The present day stellar mass of the NB is
$1.4 \pm 0.6 \times 10^9 M_\odot$ \cite{Launhardt2002}, 
which, using the results of ref.~\cite{Maraston1998}, has to be revised upwards by a factor $\sim 1.8$
to estimate the ZAMS mass of all stars formed in the NB over Galactic history;
we therefore need to rescale our reference data set
by a factor $1.25 \pm 0.37$.
Overall, we predict that $(11.5 \pm 3.4) \times 10^3$
AIC MSPs have been created in the NB out to 13.8 Gyr cosmological time.

\begin{figure}[h]
        \adjustimage{width=0.8\linewidth}{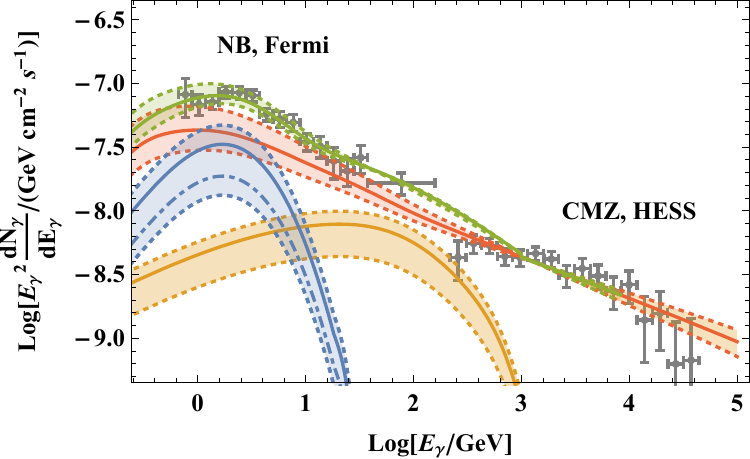}
        % \decoRule
            \caption{{\bf The $\gamma$-ray emission from the Nuclear Bulge (NB)/Central Molecular Zone (CMZ).} The spectral data points in the $\sim$ GeV regime for the NB are from ref.~\cite{Macias2019} and those in the TeV regime for the CMZ (due to the HESS instrument) are from ref.~\cite{Abdalla2018}. As previously, MSP magnetospheric emission is shown in {\bf blue} (with the dot-dashed curve showing the magnetospheric emission directly-estimated from the NB model MSP population at 13.8 Gyr and using the central values for the NB stellar mass, the fiducial binarity fraction of 70\%, etc), IC emission in {\bf yellow}, and total emission in {\bf green}, but note that we allow for a separate hadronic component (shown in {\bf red}) corresponding to $\gamma$-ray emission by cosmic ray protons and ions off gas nuclei in the CMZ spatially coincident with the NB. 
            Coloured error bands show 1-$\sigma$ errors.
        }
        \label{fig:plotNB}
\end{figure}

Given the current period and period derivatives of this model MSP population we can, as before, then
make a preliminary estimate for the aggregated prompt
emission from this population using Eq.~2 from the main text; the central value of this  is shown as the dot-dashed curve in \autoref{fig:plotNB}.
As for the case of the wider scale Galactic bulge investigated in the main text, we find that 
this initial estimate produces a peak in the spectral energy distribution at the correct $\gamma$-ray energy.
However the amplitude of this peak is more noticeably in deficit with respect to the measured spectrum and, as before, we miss the $\gtrsim 10$ GeV tail of the NB signal.
On the other hand, the measured \cite{Macias2019} luminosity of the 
GeV-band emission associated to the NB stellar template
is
$(3.9 \pm 0.5) \times 10^{36}$ erg s$^{-1}$, while our model NB MSP population's spin-down power at 13.8 Gyr is
$(8.3 \pm 3.1) \times 10^{36}$ erg s$^{-1}$.
Thus, there seems to be elegantly sufficient spin-down power to drive the observed luminosity of the NB signal and, as previously, we are motivated to model-in an additional spectral component: the IC emission off the NB lightfield by the 
CR $e^\pm$ pairs launched by the MSPs.

Before going on to blindly follow our previous approach, however, we note the following:
Aside from its distinct star formation history, a further complication encountered in modelling the 
NB's $\gamma$-ray emission is that this stellar population is 
spatially coincident \cite{Launhardt2002}
with the Central Molecular Zone (CMZ),
the dense and massive agglomeration of molecular gas surrounding the Galactic supermassive black hole out to $\sim 200$ pc radii \cite{Morris1996}.
The CMZ gas not only fuels the
ongoing star formation hosted by the NB, 
it provides target nuclei for hadronic $\gamma$-ray production by the cosmic ray population accelerated in concert with that star formation.
Indeed, the rather tight spatial correlation \cite{Aharonian2006,Abdalla2018} between 
the CMZ molecular gas column and its 
$\gamma$-ray surface brightness in the $\sim$ 1-100 TeV range demonstrate
that, at such energies, most of the $\gamma$-ray emission from the direction of the NB is actually hadronic in origin.
Furthermore,
it has long been known (e.g., \cite{Chernyakova2011}) that the GeV + TeV data covering the NB/CMZ cannot be 
well explained as due to either single power-law parent hadron or lepton distributions.
Thus, unlike the case of the wider-scale Boxy Bulge (which does not host large amounts of molecular gas target material for hadronic $\gamma$-ray production) treated in the main text, here in our spectral modelling of the NB we must account for 
hadronic emission from the CMZ 
in addition to the prompt magnetospheric and IC emission energised by the NB pulsars.

Apart from accounting for the hadronic component (with free normalisation and spectral index of the parent hadron distribution), our spectral fitting procedure for the NB is broadly similar to that described above for the Boxy Bulge: 
\begin{enumerate}
\item We allow for the prompt component to be renormalised (with respect to the preliminary estimate obtained directly from the BPS and using fiducial values for the stellar mass, binary fraction, etc.), but introduce a term into the 
$\chi^2$ to penalise for such a renormalisation.
\item We allow for the NB magnetic field to freely float, but constrain the overall fit by requiring that the predicted synchrotron emission from the CR $e^\pm$ pairs not exceed the measured (see \cite{Crocker2010} and references therein) radio emission from the region.
\item In calculating the IC emission, we adopt a interstellar radiation field estimate from GalProp \cite{Porter2008}; this has a total energy density of 15.9 eV cm$^{-3}$.
\item In distinction to our previous procedure, 
we do {\it not} allow the spectral parameters (spectral index and cut-off energies of the prompt emission and the IC-emitting CR $e^\pm$ pairs, respectively) to float; rather, we set these equal to their best-fit values found in fitting the broadband emission from the Boxy Bulge.
This procedure ensures that our overall model is physically self-consistent in the sense that the MSPs in the BB and NB are taken to
emit prompt radiation and 
launch CR $e^\pm$ pairs described by the same distributions.
\item As for the BB fitting, we introduce a term into the $\chi^2$ function that penalises the fit for any difference between the estimated spin-down power of the MSP population as derived from the BPS and the total non-thermal luminosity of the MSPs (i.e., a good fit requires that the prompt + IC + synchrotron emission more-or-less saturates the spin-down power).
\end{enumerate}

With all these in place, we arrive at the GeV-TeV $\gamma$-ray spectral fit shown by the coloured curves in \autoref{fig:plotNB}.
We find a minimum $\chi^2 = 29.8$ for $33$ (data points) - 5 (fitting parameters) = 28 (degrees of freedom), or  a reduced $\chi^2 = 1.06$, an excellent fit.
Lending us further confidence in this scenario, although it is not enforced, for the best fit we find that the ratio of prompt luminosity to overall CR $e^\pm$ pair luminosity is $0.088^{+ 0.005}_{ - 0.006}$, while the modelling of the BB returned a best fit value of $0.098^{+ 0.039}_{ - 0.034}$ for this quantity.
(Of course, the current {\it absolute} luminosity per stellar mass
differs between the BB and NB because  these systems have different star formation histories, 
so their MSP populations have different aggregate spin-down luminosities per stellar mass today.)
Other best-fit values of other NB parameters are:
magnetic field amplitude $B_{\rm NB} = 46^{+ 19}_{-7}  \ \mu$G (which is consistent with ref.~\cite{Crocker2010}),
and spectral index of CR CMZ hadron population $\alpha_{\rm had} = -2.42^{+0.05}_{-0.06}$, a physically plausible value.

\clearpage

\printbibliography[segment=\therefsegment,title={Supplementary Information References}, check=onlynew]

%The corresponding author is responsible for submitting a \href{http://www.nature.com/srep/policies/index.html#competing}{competing interests statement} on behalf of all authors of the paper. This statement must be included in the submitted article file.

\end{document}